\documentclass[aps,prc,twocolumn,floatfix,superscriptaddress,nofootinbib]{revtex4-2}

\usepackage{graphicx}
\usepackage[urlcolor=green]{hyperref}
\usepackage{amsmath,bm}
\usepackage{physics}

\usepackage{color}
\usepackage[dvipsnames]{xcolor}

\newcommand{\Kompost}{K{\o}MP{\o}ST}
\newcommand{\MUSIC}{{\sc music}}
\newcommand{\Nev}{N_\mathrm{ev}}
\newcommand{\Npts}{N_\mathrm{pts}}

\renewcommand{\dd}{{\rm d}}
\newcommand{\ee}{{\rm e}}
\newcommand{\ii}{{\rm i}}

\begin{document}

\author{Renata Krupczak} \email{rkrupczak@physik.uni-bielefeld.de}
\affiliation{Fakult\"at f\"ur Physik, Universit\"at Bielefeld, D-33615 Bielefeld, Germany}
\author{Nicolas Borghini} \email{borghini@physik.uni-bielefeld.de}
\affiliation{Fakult\"at f\"ur Physik, Universit\"at Bielefeld, D-33615 Bielefeld, Germany}
\author{Hendrik Roch} \email{Hendrik.Roch@wayne.edu}
\affiliation{Department of Physics and Astronomy, Wayne State University, Detroit, Michigan 48201, USA}

\title{Mode-by-mode evolution of Pb-Pb collisions at 5.02 TeV in a hybrid model}

\begin{abstract}
  We determine the average state and the uncorrelated modes that characterize the event-by-event fluctuations of the initial state in two typical centrality classes of Pb-Pb collisions at 5.02~TeV. 
  We find that modes in a narrow central bin are similar to those in events at fixed vanishing impact parameter, while those in a mid-peripheral centrality class are affected by the impact-parameter variation.
  We study how each fluctuation mode affects observables both in the initial state and in the final state of the collisions, at the end of a state-of-the-art boost-invariant hybrid evolution with \Kompost\ + \MUSIC\ + iSS + SMASH, and show that implementing a hadronic transport cascade in such a mode-by-mode analysis with reasonable statistical noise is costly but feasible. 
\end{abstract}

\maketitle

\section{Introduction}
\label{sec:intro}

Collisions of nuclei are a valuable tool to investigate the properties of quantum chromodynamics under extreme conditions of temperature or baryon density, in particular of the quark-gluon plasma (QGP)~\cite{Chaudhuri:2012yt,Elfner:2022iae}. 
To make progress, theoretical ideas for the successive stages of the evolution have to be implemented in numerical simulations of the collisions, whose observable predictions are tested against the measurements obtained in experiments. 
While significant developments have been made in the past two decades, however, there are still open questions.

One of the largest sources of uncertainty in these simulations arises from the modeling of the initial state of the evolution, for which a large number of models have been developed~\cite{Drescher:2006ca,Miller:2007ri,Broniowski:2007nz,Albacete:2014fwa,Loizides:2014vua,Moreland:2014oya,Niemi:2015qia,Shen:2017bsr,Giacalone:2019kgg,Soeder:2023vdn,Garcia-Montero:2023gex,Kuha:2024kmq}. 
Indeed, it is well-established that final observables are highly sensitive to the initial conditions~\cite{Socolowski:2004hw,Drescher:2006ca,Broniowski:2007nz,Broniowski:2007ft,Schenke:2012wb,Schenke:2012hg,Schenke:2012wb,Paatelainen:2012at,Liu:2015nwa,Shen:2020jwv,NunesdaSilva:2020bfs, Shen:2022oyg}. 
This dependence affects the determination of QGP properties, in particular its transport coefficients, from statistical comparisons between model predictions and experimental data~\cite{Novak:2013bqa,Pratt:2015zsa,Bernhard:2015hxa,Bernhard:2016tnd,Auvinen:2017fjw,Bernhard:2019bmu,Moreland:2018gsh,Nijs:2020ors,JETSCAPE:2020shq,Nijs:2020roc,JETSCAPE:2020mzn,JETSCAPE:2020avt,JETSCAPE:2021ehl,Parkkila:2021yha,Parkkila:2021tqq,Xie:2022ght,Liyanage:2023nds,Shen:2023awv,Heffernan:2023gye,Heffernan:2023utr,Jahan:2024wpj,JETSCAPE:2024cqe,Domingues:2024pom,Gotz:2025wnv,Jahan:2025cbp}.

An avenue to make progress is to try and relate one-to-one geometrical features in the initial state and specific final-state observables. 
In the idealized case of a linear evolution from initial to final state, an analysis of measured observables in terms of principal components~\cite{Bhalerao:2014mua,Mazeliauskas:2015vea,Mazeliauskas:2015efa} would then allow to reconstruct the initial condition. 
Paralleling this idea, the event-by-event variations of the initial state have been described as the superposition of linearly propagating (small) fluctuations about an average  profile~\cite{Floerchinger:2013rya,Floerchinger:2013vua,Floerchinger:2013hza,Floerchinger:2013tya,Floerchinger:2014fta,Floerchinger:2018pje}, using a pre-determined two-dimensional basis to characterize the fluctuations~\cite{Coleman-Smith:2012kbb}.

It is however clear that the ideal of a linear response is at best approximately realized for a few choice observables, as for example the relationship between the spatial eccentricities and anisotropic-flow coefficients in central events. 
More generally, the evolution will lead to nonlinearities, such that the description in terms of an average event and fluctuation modes may quickly become intractable. 
To partly remedy this issue, a decomposition based on uncorrelated fluctuation modes has been introduced~\cite{Borghini:2022iym}, which greatly simplifies the treatment at least at quadratic order in the modes. 
While the original study focused on collisions at fixed impact parameter value, here we apply the framework to events in centrality classes, in which the impact parameter may vary, introducing an extra source of initial-state fluctuations besides those due to the random positions of nucleons in the colliding nuclei.
Furthermore, we extend the dynamical stage to include a hadronic afterburner after the fluid-dynamical evolution. 

We begin by reviewing the principle of the mode decomposition and of the mode-by-mode evolution in Sec.~\ref{sec:theory}.
We then specify the system under study, the models used to describe its evolution, and the numerical implementation of the method in Sec.~\ref{sec:numerical}.
The results are presented in Sec.~\ref{sec:results}, starting with a discussion of the fluctuation modes and continuing with their impact on initial-state characteristics and final-state observables. 
Further results are presented in the Appendices. 
Eventually, the findings are summarized in  Sec.~\ref{sec:conclusions}.

\section{Theoretical framework}
\label{sec:theory}

In this section we briefly review a number of elements of the mode decomposition of fluctuating initial states introduced in Ref.~\cite{Borghini:2022iym}: 
We first recall in Sec.~\ref{ss:modes_theory} the principle of the decomposition of initial-state profiles into an average state and uncorrelated fluctuation modes. 
Then we define the response coefficients that characterize the influence of individual modes on observables (Sec.~\ref{ss:response_theory}).

\subsection{Initial-state fluctuations and their mode decomposition}
\label{ss:modes_theory}

Starting from an initial-state model, an ensemble of $\Nev$ initial profiles $\Phi^{(i)}(\bm{x})$ will in general not be uniform. 
This variation can be interpreted as ``fluctuations'' around an average initial state $\Bar{\Psi}(\bm{x})$:
\begin{equation}
\label{eq:Phi=Psi+dPhi}
\Phi^{(i)}(\boldsymbol{x}) = \Bar{\Psi}(\bm{x}) + \delta\Phi^{(i)}(\boldsymbol{x}).
\end{equation}
Since an ``average over events'', which throughout this paper will be denoted by angular brackets, is generally an arithmetic mean, this is also the natural choice for the average initial state:
\begin{equation}
\Bar{\Psi}(\bm{x}) \equiv \big\langle\Phi^{(i)}(\bm{x})\big\rangle \equiv 
\frac{1}{\Nev} \sum_{i=1}^{\Nev} \Phi^{(i)}(\bm{x}).
\label{eq:avg}
\end{equation}

In Ref.~\cite{Borghini:2022iym} it was shown that the spectral decomposition of the autocorrelation function of fluctuations
\begin{align}
\rho(\bm{x},\bm{y}) & \equiv 
  \big\langle\delta\Phi^{(i)}(\bm{x})\, \delta\Phi^{(i)}(\bm{y})\big\rangle \cr 
  & =\frac{1}{\Nev} \sum_{i=1}^{\Nev} \Phi^{(i)}(\bm{x}) \Phi^{(i)}(\bm{y}) - \Bar{\Psi}(\bm{x})\Bar{\Psi}(\bm{y})
  \label{eq:rho(x,y)_def}
\end{align}
provides eigenvalues $\{\lambda_l\}$ and a basis of orthogonal uncorrelated ``fluctuation modes'' $\{\Psi_l(\bm{x})\}$ for the fluctuations $\delta\Phi^{(i)}(\bm{x})$. 
That is, each initial state $\Phi^{(i)}(\bm{x})$ can be decomposed in the form
\begin{equation}
\Phi^{(i)}(\boldsymbol{x}) = \Bar{\Psi}(\boldsymbol{x}) + \sum_l c_l^{(i)} \Psi_l(\boldsymbol{x}),
\label{eq:decomposition}
\end{equation}
where the coefficients $\{c_l^{(i)}\}$ in the linear combination of modes are such that 
\begin{align}
\big\langle c_l^{(i)} \big\rangle &= 0, \label{eq:<c_l>=0}\\
\big\langle c_l^{(i)} c_{l'}^{(i)} \big\rangle &= 0\text{ for }l\neq l'.  \label{eq:<c_l.c_l'>_1}
\end{align}
With an appropriate normalization of the fluctuation modes, namely $\norm{\Psi_l} = \sqrt{\lambda_l}$, with the norm that of square-integrable functions, one can ensure that the expansion coefficients have a unit variance: 
\begin{equation}
\Big\langle \big(c_l^{(i)}\big)^2 \Big\rangle = 1.
\label{eq:<c_l.c_l'>_2}
\end{equation}

In a numerical implementation, the initial profiles $\Phi^{(i)}(\bm{x})$ are represented by vectors (one-dimensional arrays), which we also denote by $\Phi^{(i)}$, with a dimension equal to the number $\Npts$ of discretization points of space in case $\Phi^{(i)}$ stands for a single scalar quantity like the energy density.%
\footnote{If an initial profile consists of a ``thermodynamic'' quantity, like the energy density, together with $N_{\rm ch}$ densities of conserved charges, then the number of entries of a vector $\Phi^{(i)}$ is $(N_{\rm ch}+1)\Npts$.}
Computing the average initial state $\bar{\Psi}$ is then straightforward from the definition~\eqref{eq:avg}.
The modes $\{\Psi_l\}$ and the corresponding eigenvalues $\{\lambda_l\}$ are obtained by diagonalizing the $\Npts\times\Npts$ matrix
\begin{equation}
\label{eq:rho}
\rho \equiv \frac{1}{N_{\mathrm{ev}}} \sum_{i=1}^{N_\mathrm{ev}} \Phi^{(i)} \Phi^{(i)\mathsf{T}} - \Bar{\Psi}\Bar{\Psi}^{\mathsf{T}},
\end{equation}
which is the discretized version of Eq.~\eqref{eq:rho(x,y)_def}.
The fluctuation mode $\Psi_l$ is namely an eigenvector (with norm $\sqrt{\lambda_l}$) for the eigenvalue $\lambda_l$ of this matrix. 
Once the modes have been determined, the expansion coefficients $\{c_l^{(i)}\}$ for a given initial state $\Phi^{(i)}$ are readily determined by projecting $\delta\Phi^{(i)}$ on the $\{\Psi_l\}$.

From now on we shall mostly omit the superscript $(i)$ labeling events.

\subsection{Mode-by-mode response of observables}
\label{ss:response_theory}

The interest of the decomposition in uncorrelated modes is that it allows a faster computation of the event-by-event fluctuations of final-state observables, as we now explain.

Let $O_\alpha$ denote an observable. 
In practice, it can be an initial-state quantity, that only depends on the initial profiles $\{\Phi^{(i)}\}$, or a ``final-state'' observable, after some dynamical evolution stage. 
For each initial state, the value of the observable is $O_\alpha(\Phi^{(i)})$, where one can replace $\Phi^{(i)}$ by its decomposition~\eqref{eq:decomposition}.
Provided the observable is well enough behaved, one can write down a Taylor expansion for this value:
\begin{equation}
O_\alpha \big(\Phi^{(i)}\big) = 
 \Bar{O}_\alpha + \sum_l L_{\alpha, l} c_l^{(i)} + 
 \frac{1}{2} \sum_{l, l'} Q_{\alpha, ll'} c_l^{(i)} c_{l'}^{(i)} + {\cal O}(c_l^3),
\label{eq:observables}
\end{equation}
where we introduced the notations 
\begin{equation}
\Bar{O}_\alpha \equiv O_\alpha(\Bar{\Psi})
\label{eq:def_O-bar}
\end{equation}
for the value of the observable computed for the average initial state, and
\begin{equation}
L_{\alpha, l} \equiv \frac{\partial O_\alpha}{\partial c_l}\bigg|_{\Bar{\Psi}}
\label{eq:def_L_l}
\end{equation}
and
\begin{equation}
Q_{\alpha, ll'} \equiv \frac{\partial^2 O_\alpha}{\partial c_{l\,} \partial c_{l'}}\bigg|_{\Bar{\Psi}}
	\label{eq:def_Q_ll'}
\end{equation}
for the first and second derivatives of the observable, evaluated at $\bar{\Psi}$, respectively. 
We shall refer to the $L_{\alpha,l}$ and $Q_{\alpha, ll'}$ as the linear and quadratic-response coefficients, respectively. 

Since the expansion coefficients $\{c_l\}$ are by construction of order 1, truncating Eq.~\eqref{eq:observables} at order $c_l^2$ may already provide a good approximation of the value of the observable. 
From Eqs.~\eqref{eq:observables}--\eqref{eq:def_Q_ll'} the event-average of an observable is then readily computed. 
Under consideration of Eqs.~\eqref{eq:<c_l>=0}--\eqref{eq:<c_l.c_l'>_2}, one has
\begin{equation}
\label{eq:<O_a>}
\expval{O_\alpha} \equiv 
  \frac{1}{N_{\mathrm{ev}}} \sum_{i=1}^{N_\mathrm{ev}} O_\alpha \big(\Phi^{(i)}\big) \simeq 
  \bar{O}_\alpha + \frac{1}{2}\sum_l Q_{\alpha, ll},
\end{equation}
where only the diagonal quadratic-response coefficients enter, thanks to Eqs.~\eqref{eq:<c_l>=0} and~\eqref{eq:<c_l.c_l'>_1}. 
In turn, the covariance of two observables $O_\alpha$ and $O_\beta$ is
\begin{equation}
\label{eq:<O_a.O_b>}
\expval{\big(O_\alpha-\expval{O_\alpha}\!\big)\big(O_\beta-\expval{O_\beta}\!\big)} \simeq 
\sum_l L_{\alpha, l} L_{\beta, l},
\end{equation}
up to terms of order $c_l^3$. 
This time, only the linear-response coefficients are involved. 

Equations~\eqref{eq:<O_a>} and \eqref{eq:<O_a.O_b>} mean that the average values and (co)variances of observables, defined over an ensemble of events, can be approximated by mode-by-mode expressions, which rely on that of the response coefficients~\eqref{eq:def_L_l}--\eqref{eq:def_Q_ll'}. 
For final-state observables, this mode-by-mode calculation is significantly less time-consuming. 
One only needs to compute numerically the first and second derivatives~\eqref{eq:def_L_l} and \eqref{eq:def_Q_ll'} (with $l'=l$). 
For instance, denoting by $\xi$ a small number --- example values will be given in Sec.~\ref{sec:results} --- and by $\Psi^\pm_l \equiv \Bar{\Psi} \pm \xi \Psi_l$ the mock initial states that result from adding or subtracting the single-mode fluctuation $\delta\Phi_l = \xi\Psi_l$ to the average initial state $\bar{\Psi}$, one can compute the values of a specific observable $O_\alpha$ for these two states: $O^\pm_{\alpha, l} \equiv O_\alpha(\Psi^\pm_l) = O_\alpha(\Bar{\Psi} \pm \xi \Psi_l)$.
With their help, one can then obtain the response coefficients as
\begin{equation}
L_{\alpha, l} = \frac{O^+_{\alpha, l} - O^-_{\alpha, l}}{2 \xi}
\quad\text{and}\quad
Q_{\alpha, ll} = \frac{O^+_{\alpha, l} + O^-_{\alpha, l} - 2\Bar{O}_\alpha}{\xi^2},
\label{eq:L&Q_num}
\end{equation}
where $\Bar{O}_\alpha$, Eq.~\eqref{eq:def_O-bar}, is the same for all modes.

\section{Numerical approach}
\label{sec:numerical}

In this study the profiles $\{\Phi^{(i)}\}$ are energy densities for the initial state at a time $\tau_0$ of longitudinally boost invariant Pb-Pb collisions at 5.02~TeV. 
More precisely, these are discretized profiles on a 2D grid with $\Npts = 192\times 192$ points with a physical spacing of 0.11~fm.
The $\{\Phi^{(i)}\}$ are generated with a Monte Carlo Glauber model (MC Glauber)~\cite{Miller:2007ri,dEnterria:2020dwq} that samples the position of nucleons inside each Pb nucleus based on a Woods--Saxon distribution with half-density radius $R = 6.62$~fm and diffusivity $a = 0.546$~fm, implementing a minimum distance of 0.4~fm between two nucleons. 
The occurrence of nucleon-nucleon collisions relies on a geometric criterium: 
two nucleons collide when their distance in the transverse plane is smaller than or equal to $(\sigma^\mathrm{NN}_\mathrm{inel}/\pi)^{1/2}$, with $\sigma^\mathrm{NN}_\mathrm{inel} = 67.6$~mb. 
After defining the local numbers of participants ($N_\mathrm{part}$) and binary collisions ($N_\mathrm{coll}$), the energy density at midrapidity is assumed to be of the form%
\begin{equation}
\label{eq:e(x)}
 e(\bm{x}) \propto (1-\alpha) \frac{N_\mathrm{part}(\bm{x})}{2} + \alpha N_\mathrm{coll}(\bm{x})
\end{equation}
with $\alpha = 0.2$~\cite{dEnterria:2020dwq}. 
This energy density is deposited as Gaussian distributions with a width of $0.4$~fm and centered at the nucleon positions (for the $N_\mathrm{part}$-component) or at the halfway point between the centers of two colliding nucleons (for the $N_\mathrm{coll}$-component). 
The resulting profile is then re-centered --- such that the center-of-mass of $e(\bm{x})$ coincides with the center of the grid --- and normalized with a multiplicative $K$-factor to produce the correct final charged-hadron multiplicity $\dd N_{\rm ch}/\dd\eta$ per unit pseudorapidity.
To quickly compute the latter, we relate it to the initial energy density via the estimator formula introduced in Ref.~\cite{Giacalone:2019ldn}, which states that the final multiplicity is proportional to the integral over the transverse plane of $e(\bm{x})^{2/3}$. 
In turn, the estimator formula needs to be calibrated to yield the same multiplicity as dynamical calculations starting from the same initial $e(\bm{x})$, which can be done using only a small number of events. 
Demanding that the results match the multiplicity values reported by the ALICE Collaboration~\cite{ALICE:2015juo}, we find that the $K$-factor multiplying Eq.~\eqref{eq:e(x)} should be equal to $37.53$~GeV$/_{}$fm$^2$.

Hereafter we present results for Pb-Pb collisions at $5.02$~TeV in two centrality classes: 0--2.5\% and 30--40\%. 
Note that the impact parameter of the collisions (defined by the centers of the two Woods--Saxon distributions), keeps a fixed direction, which defines the $x$-axis, although its value can vary. 
For each centrality bin, we produced $2^{21}$ energy-density profiles, from which the corresponding average state $\Bar{\Psi}$ and $\rho$-matrix [Eq.~\eqref{eq:rho}] are determined.
Diagonalizing the latter to obtain its eigenvalues $\{\lambda_l\}$ and eigenvectors\footnote{Since $\rho$ is a square matrix of size $192^2 \cross 192^2$, this step involves significant computational costs in terms of memory and time.} yields the set of fluctuation modes $\{\Psi_l\}$. 
While the theoretical decomposition in Eq.~\eqref{eq:decomposition} in principle involves an infinite number of modes, the numerical implementation is limited by the grid size. 
However, as will be demonstrated in Sec.~\ref{sec:results}, the first few modes capture the largest contributions to the random profiles, which is sufficient to extract the main features of the fluctuations. 
%With this information, the initial analysis of the modes can be performed, and the results are shown in Sec. \ref{sec:IS}.

For the subsequent dynamical evolution of the initial states~\cite{Roch-Krupczak2024}, a full energy-momentum tensor is determined at each point of the transverse plane from the energy density $e(\bm{x})$ provided by the MC Glauber model, in the form $T^{\mu\nu}(\bm{x}) = \mathrm{diag}(e(\bm{x}), e(\bm{x})/2, e(\bm{x})/2, 0)$. 
Note that we use a different grid for this dynamical evolution, with a total box size of 30~fm to ensure that the expanding system does not come into contact with the edges, and a spacing of 0.1~fm (obtained by interpolating the values on the initial grid).

Starting from the initial state at $\tau_0 = 0.2$~fm, we use \Kompost~\cite{Kurkela:2018vqr} for the first, pre-equilibrium stage of the evolution of $T^{\mu\nu}(\bm{x})$. 
\Kompost\ is run with a constant specific shear viscosity $\eta/s = 0.16$ and considering only energy perturbations, until $\tau_{\mathrm{hydro}} = 1.0$~fm. 
At that time, the stress tensor is isotropic enough to warrant the application of dissipative fluid dynamics for the evolution. 

To switch from \Kompost\ to the hydrodynamic code \MUSIC~\cite{Schenke:2010nt, Schenke:2010rr, Paquet:2015lta}, we use the traditional Landau matching procedure based on the energy densities, rather than the entropy-matching recipe introduced in Ref.~\cite{Borghini:2024kll} to mitigate the mismatch between the equations of state (EoS) in the two systems. 
Indeed, we run \MUSIC\ with the EoS from the HotQCD collaboration~\cite{HotQCD:2014kol}, matched with the SMASH resonance gas at lower temperatures, which differs from the conformal EoS within \Kompost. 
In \MUSIC, we use the same constant specific shear viscosity $\eta/s = 0.16$ as in \Kompost\ and the temperature-dependent parametrization of the bulk density over entropy ratio $\zeta/s$ introduced in Ref.~\cite{Bernhard:2018hnz}. 
The system is evolved with \MUSIC\ until each cell reaches the particlization temperature of $T_{\mathrm{f.o.}} = 155$~MeV, at which a freeze-out hypersurface is determined. 

After producing this hypersurface, we explore two different setups for the production of the final state. 
On the one hand, we use the freeze-out procedure implemented in the \MUSIC\ code, which produces the momentum distributions of the various particle species. 
These are then allowed to decay but do not rescatter. 
On the other hand, we employ the iSS package~\cite{Shen:2014vra} to convert the continuous degrees of freedom into hadrons. 
The latter are then input into SMASH~\cite{SMASH:2016zqf}, which lets them scatter and decay, and is responsible for producing the final list of particles. 
More details on either approach will be given in Sec.~\ref{sec:FS}.

\section{Results}
\label{sec:results}

In this section we present the results of our mode-by-mode analysis. 
We begin with displaying in Sec.~\ref{sec:mode-results} the output of the mode decomposition for both centrality classes, namely the corresponding average initial state and fluctuation modes. 
In Sec.~\ref{sec:IS} we introduce initial-state characteristics, and discuss how they are affected by the modes. 
We finally turn to results on final-state observables in Sec~\ref{sec:FS}.

\subsection{Average initial state and fluctuation modes}
\label{sec:mode-results}

\begin{figure}
    \includegraphics[width=\linewidth]{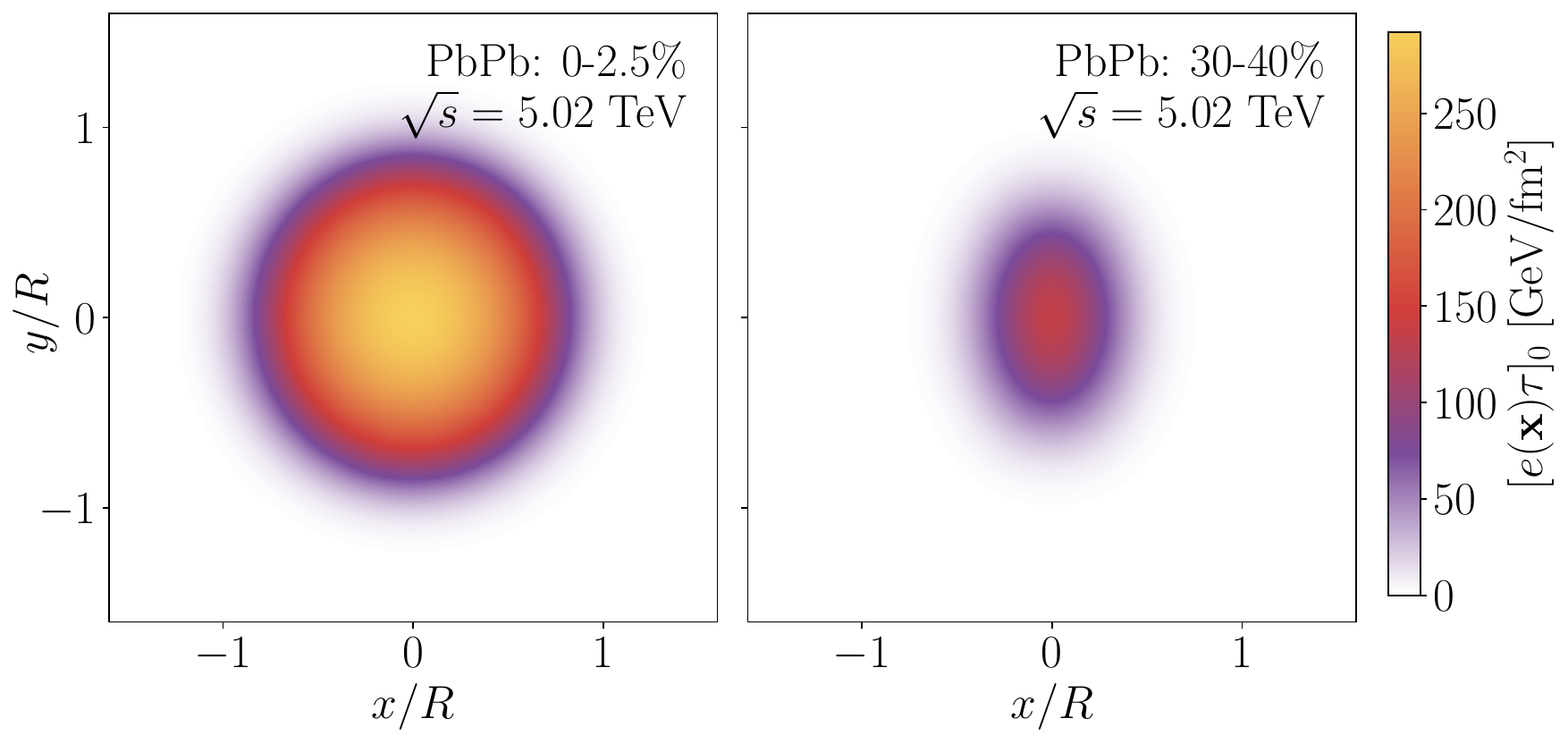}\vspace{-2mm}
    \caption{Average initial state $\Bar{\Psi}$ for events in the 0--2.5\% (left) and 30--40\% (right) centrality bins. 
    In each bin the average is computed over $\Nev = 2^{21}$ initial states.
    Both axes are in units of the half-density radius $R=6.62$~fm.}
\label{fig:avg}
\end{figure}

Starting from the $\Nev = 2^{21}$ energy-density profiles generated in each centrality bin, the average initial state $\bar{\Psi}$ computed using Eq.~\eqref{eq:avg} is shown in Fig.~\ref{fig:avg} for central (0--2.5\%) events on the left and mid-peripheral ones (30--40\%) on the right.\footnote{The robustness of the results reported in this section was checked using simulations on a coarser grid with $128\times 128$ points, corresponding to a spacing of 0.16~fm.}
As could be anticipated, these average states are smooth, in contrast to the individual initial states from random Pb-Pb collisions. 
Since the impact-parameter direction is fixed, the shapes of the two profiles visibly differ: $\bar{\Psi}$ is more circular for the central events, with an eccentricity of $\varepsilon_2 = 0.0174$ (the quadrangularity $\varepsilon_4$ is of order $10^{-4}$), and more almond-shaped for the non-central events, with $\varepsilon_2 = 0.3214$ and $\varepsilon_4 = 0.1132$ (with a 4th-order symmetry plane angle rotated by $\Phi_4=\pi/4$ with respect to the impact-parameter direction.)
One also sees that the energy density reaches higher values in central collisions, in which more energy overall is deposited, which was to be expected. 

\begin{figure}
	\includegraphics[width=\linewidth]{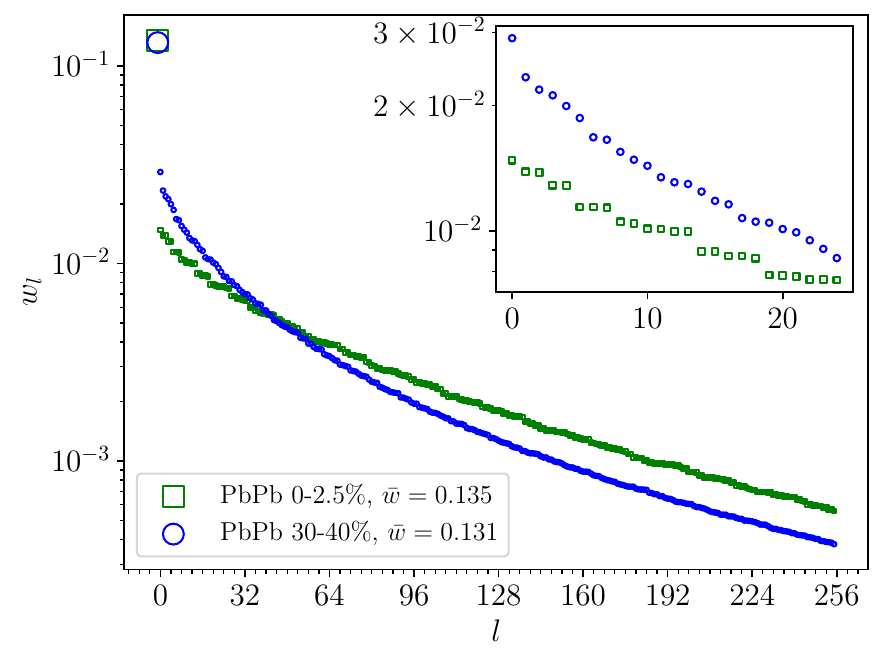}\vspace{-3mm}
	\caption{Relative weights $\{w_l\}$ of the first 256 fluctuation modes and of the average initial state ($\bar{w}$, larger symbols at $l = -1$) for each centrality class.}
	\label{fig:weights}
\end{figure}

After the average state, one can determine the fluctuation modes $\{\Psi_l\}$ following the procedure outlined in Sec.~\ref{sec:theory}. 
Since the expansion coefficients $\{c_l\}$ in Eq.~\eqref{eq:decomposition} are by construction of order 1,\footnote{In Appendix~\ref{apx:cl} we present a few results on the statistics of the $\{c_l\}$ calculated from a sample of random events, in particular to illustrate Eqs.~\eqref{eq:<c_l>=0}--\eqref{eq:<c_l.c_l'>_2}.} the typical contribution of a given mode $\Psi_l$ to a random initial state is given by its norm $\norm{\Psi_l}$. 
To compare these contributions of the various modes with each other, and with the contribution from the average initial state $\bar{\Psi}$, we introduced the relative weights~\cite{Borghini:2022iym}
\begin{equation}
w_l \equiv \frac{\norm{\Psi_l}}{\sum_l \norm{\Psi_l} + \norm{\Bar{\Psi}}}
\quad\text{and}\quad
\bar{w} \equiv \frac{\norm{\Bar{\Psi}}}{\sum_l \norm{\Psi_l} + \norm{\Bar{\Psi}}},
\end{equation}
where the latter measures the relative importance of $\bar{\Psi}$. 
Note that the sum in the denominator extends over 6900 modes, even though we only show results for a small subset.\footnote{In Ref.~\cite{Borghini:2022iym} the corresponding sum was computed using all $\Npts$ modes. 
  We now use another, quicker diagonalization routine, that only yields the larger eigenvalues (and corresponding eigenstates), which explains why we no longer sum over all modes. 
  Yet we find that $w_l$ is smaller than $10^{-8}$ (and still decreasing) for $l\gtrsim 4000$, so that the modes not taken into account actually contribute very little.} 
Throughout the paper, the index $l$ labeling the modes is such that they are ordered by decreasing relative weight $w_l$, with $l$ starting at~0. 
Thus, when referring to the ``first modes'', we mean those with the largest contributions to a typical random initial state. 

In Fig.~\ref{fig:weights} we show the relative weights of the average initial state and of the 256 first fluctuation modes for the two centrality classes. 
For both central and non-central collisions the relative weight $\bar{w}$ of the average initial state is about 13\%, similar to the values found for MC Glauber initial states in collisions at fixed impact parameter~\cite{Borghini:2022iym}.
Other trends in Fig.~\ref{fig:weights} are very similar to those reported in Ref.~\cite{Borghini:2022iym}.
Thus, the relative weights of the first modes are larger, ranging up to 3\%, in non-central events than in central collisions, where $w_l$ stays below 2\%. 
Also, the spectrum of $w_l$ values is flatter in central collisions than in non-central ones. 
Both behaviors parallel the difference between events with $b=0$ and those with $b=9$~fm. 
In Ref.~\cite{Borghini:2024ekn}, it was argued that the slope of the spectrum of $w_l$ values is affected by correlations between fluctuations at different positions in the transverse plane, with larger correlations possibly leading to a steeper spectrum. 
While a more systematic investigation is still needed, this may suggest that the point-to-point correlations in the initial states of non-central events are more important than in central ones.

\begin{figure*}
	\includegraphics[width=.85\linewidth]{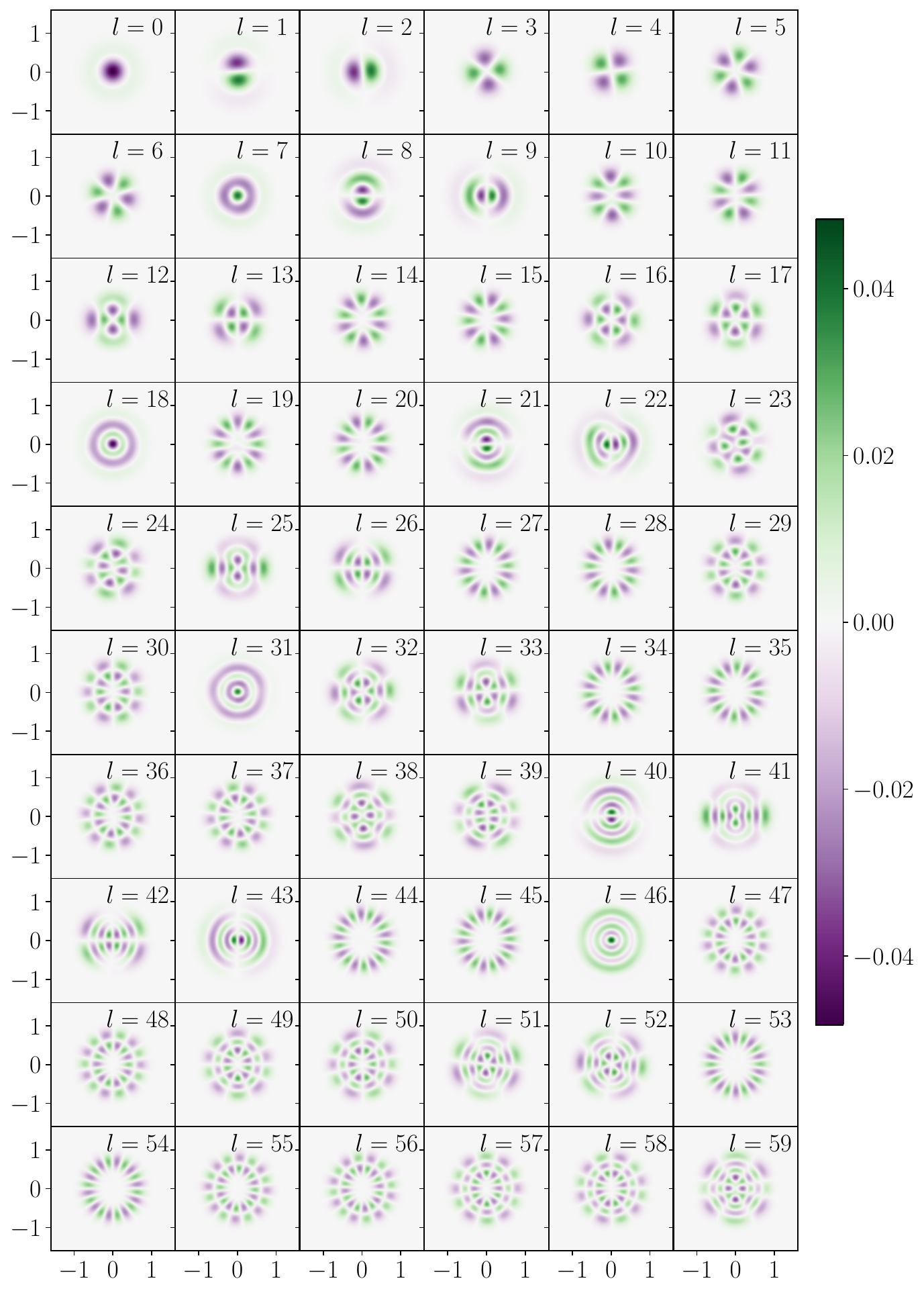}\vspace{-3mm}
	\caption{Normalized transverse profile of the first $60$ modes for central events. Both axes are in units of the half-density radius $R=6.62$~fm.}
	\label{fig:modes0}
\end{figure*}

\begin{figure*}
	\includegraphics[width=.85\linewidth]{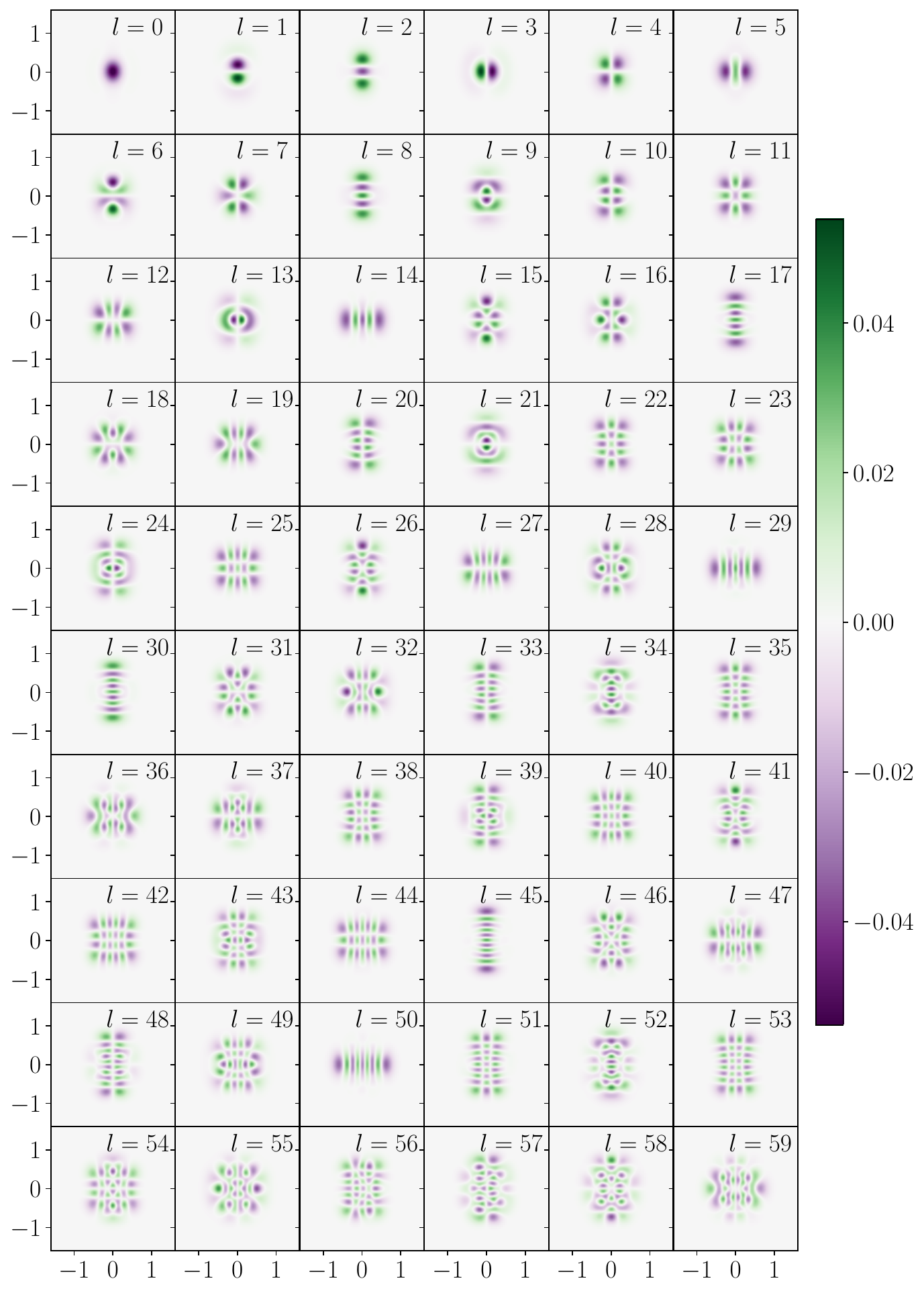}\vspace{-3mm}
	\caption{Normalized transverse profile of the first $60$ modes for events in the 30--40\% centrality bin. Both axes are in units of the half-density radius $R=6.62$~fm.}
	\label{fig:modes30}
\end{figure*}

Despite the difference in steepness, the relative weight decreases rather quickly in either centrality class, which suggests that the first modes will indeed be the dominant ones and contribute most to observables. 
Focusing on the relative weights of the first 25 modes (see inset in Fig.~\ref{fig:weights}), a marked difference appears between central and non-central events. 
For the former, there tends to be pairs of (quasi)degenerate weights, like $w_1$ and $w_2$ or $w_3$ and $w_4$, which are readily explained when one looks at the profiles of the respective modes, which we shall discuss hereafter. 
In contrast, no such degeneracy --- or only an accidental one --- is observed in initial states of non-central collisions. 
In Ref.~\cite{Borghini:2022iym}, the degeneracy was observed both in events at vanishing impact parameter and in those at fixed $b=9$~fm. 
Accordingly, we may safely ascribe the lifting in degeneracy for the events in the 30--40\% centrality class to the variation of the impact parameter, which acts as an extra ``source'' of fluctuations besides the randomness in the position where individual nucleon-nucleon collisions take place. 

The transverse profiles of the first 60 fluctuation modes $\{\Psi_l\}$ are shown in Fig.~\ref{fig:modes0} for central events and in Fig.~\ref{fig:modes30} for events in the 30--40\% centrality bin. 
Since the norm of $\Psi_l$, or equivalently the relative weight $w_l$, decreases by a factor of approximately 3.7 between $\Psi_0$ and $\Psi_{59}$ in the central bin --- in the 30--40\% centrality class, the norms differ by a factor 7.8 ---, we do not display the modes themselves, but rather normalized eigenvectors, to allow a better comparison.\footnote{Accordingly, the scale of the color code has no physical meaning, and in particular it depends on the grid step.}
In Appendix~\ref{apx:modes2} we also display the next 60 normalized eigenvectors, corresponding to the modes $\{\Psi_l\}$ with $60\leq l\leq 119$, for both centrality bins.

A first cursory glance comparing the eigenvectors in the two centrality classes reveals a few similarities, to which we will come shortly, but also one striking difference:
While the fluctuation modes in central collisions have a roughly round profile, as does the corresponding average initial state, this is not true of the majority of eigenvectors shown in Fig.~\ref{fig:modes30}. 
Indeed, some of latter are even of almost rectangular shape with one side parallel to the $x$-axis, i.e.\ along the impact-parameter direction. 
The difference naturally comes from the fact that events in the 30--40\% centrality bin mostly have a significant impact parameter of about 7--9~fm, while central events have an almost vanishing impact parameter.

Besides this difference, there are however similar qualitative features across the two centralities, that were already present in the study at fixed impact parameter~\cite{Borghini:2022iym}.
First, all modes comprise regions with both positive and negative energy density --- or, rather, positive and negative contributions to the energy density profile of a random initial state.\footnote{Since the negative of an eigenvector is also an eigenvector, the fact that, for instance, the eigenvectors with $l=0$ in Figs.~\ref{fig:modes0} and~\ref{fig:modes30} are negative in the center has no physical meaning, but is the random outcome of the diagonalization routine.} 
Since each fluctuation mode only yields a small contribution to any random event, which also involves the much larger contribution from the average initial state $\bar{\Psi}$, see Eq.~\eqref{eq:decomposition}, the regions with a negative energy density contribution (when accounting also for the expansion coefficient $c_l$) only signal that the mode locally decreases the energy density compared to $\bar{\Psi}$ --- and increases it locally elsewhere ---, but overall the energy density remains non-negative everywhere. 

Secondly, the various modes exhibit different azimuthal structures with approximate circular or elliptic symmetry, or with clear dipoles, quadrupoles, octupoles, and so on. 
In central collisions, the eigenvectors with such a multipolar structure come in pairs related by a simple rotation: for instance, the eigenvectors with $l=1$ and $l=2$ in Fig.~\ref{fig:modes0} are rotated by $90^{\rm o}$, those with $l=3$ and $l=4$ are rotated by $45^{\rm o}$, those with $l=5$ and $l=6$ are rotated by $30^{\rm o}$. 
These pairs of related eigenvectors are precisely those with (quasi)degenerate relative weights $w_l$. 
In the 30-40\% centrality bin (Fig.~\ref{fig:modes30}), one recognizes fewer such pairs of eigenvectors (approximately) rotated with respect to each other, for instance $l=1$ and $l=3$ or possibly $l=22$ and $l=25$, and they do not have degenerate weights: the inset in Fig.~\ref{fig:weights} clearly shows $w_1>w_3$.
 
 \begin{figure*}
 	\includegraphics[width=\linewidth]{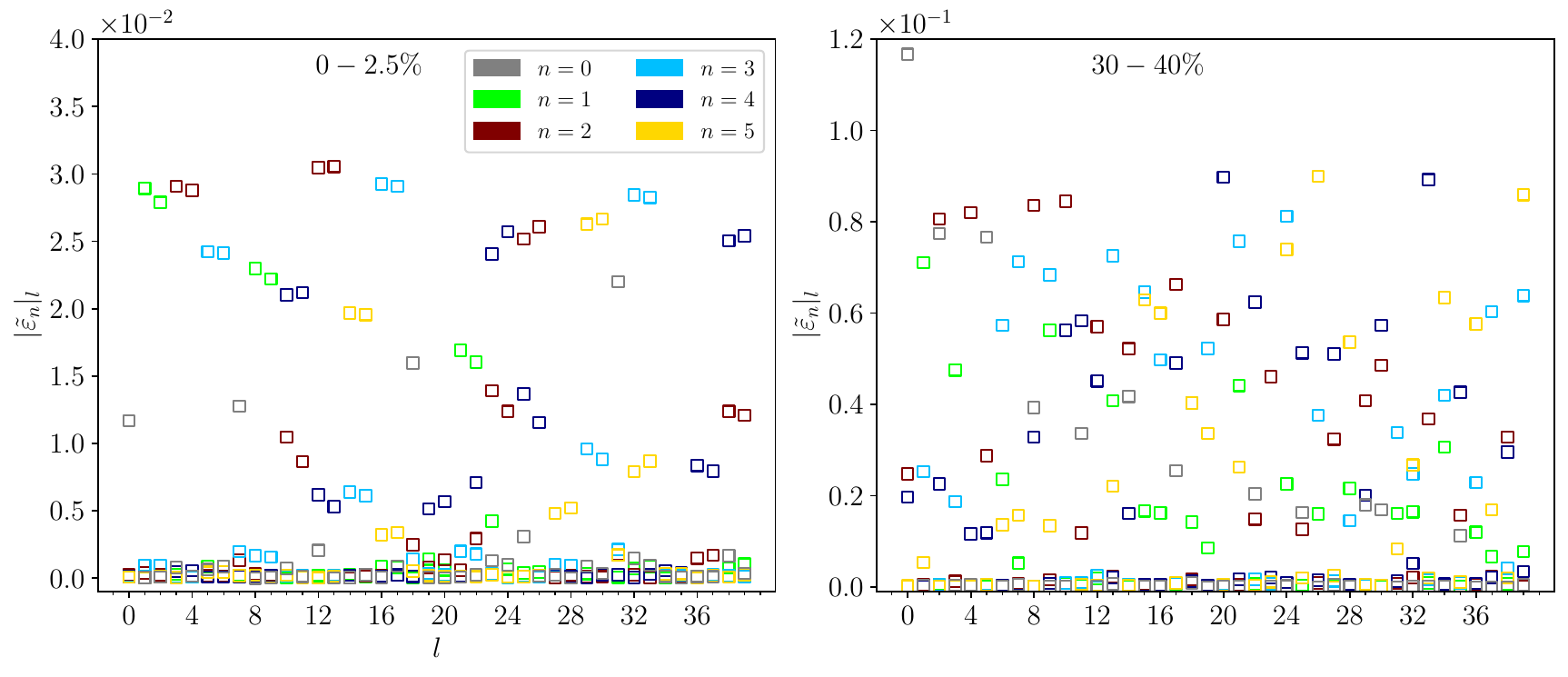}\vspace{-3mm}
 	\caption{Relative energy content ($|\tilde{\varepsilon}_0|_l$) and spatial anisotropies $|\tilde{\varepsilon}_n|_l$ [see Eqns.~\eqref{eq:mode_eccentricities1}--\eqref{eq:mode_eccentricities2}] with $1\leq n\leq 5$ of the first 40 fluctuation modes for events in the 0--2.5\% (left) and the 30-40\% (right) centrality class.}
 	\label{fig:mode_eccentricities}
 \end{figure*}
 
To go beyond the visual impression, we introduce a few quantities to characterize the modes $\{\Psi_l\}$ quantitatively. 
Introducing a centered system of polar coordinates $(r,\theta)$, we define the dimensionless quantities
\begin{equation}
|\tilde{\varepsilon}_n|_l \equiv 
	\frac{\bigg|\displaystyle\int\! r^n \ee^{\ii n\theta}\Psi_l(r,\theta)\,r\,\dd r\,\dd\theta\bigg|}%
	{\displaystyle \int\! r^n\bar{\Psi}(r,\theta)\,r\,\dd r\,\dd\theta}
	\quad\text{for }n\neq 1
	\label{eq:mode_eccentricities1}
\end{equation}
and
\begin{equation}
|\tilde{\varepsilon}_1|_l \equiv 
	\frac{\bigg|\displaystyle\int\!r^3 \ee^{\ii\theta} \Psi_l(r,\theta)\,r\,\dd r\,\dd\theta\bigg|}%
	{\displaystyle \int\!r^3 \bar{\Psi}(r,\theta)\,r\,\dd r\,\dd\theta}
	\quad\text{for }n=1. \label{eq:mode_eccentricities2}
\end{equation}
One readily sees that $|\tilde{\varepsilon}_0|_l$ represents the absolute value of the energy (per unit space-time rapidity) of mode $l$ divided by that of the average initial state. 
In turn, $|\tilde{\varepsilon}_n|_l$ for $n\geq 1$ is related to the modulus of the $n$-th order eccentricity --- with the important difference that we divide by a moment of $\bar{\Psi}$, not of $\Psi_l$ itself, to make sure that the denominator is always nonzero.
These quantities with $0\leq n\leq 5$ are plotted in Fig.~\ref{fig:mode_eccentricities} for the first 40 fluctuation modes in both centrality classes. 
In central collisions, only very few of these modes ($l=0$, 7, 18, 31) have a sizable energy content, of order 1--2\% of that of the average initial state. 
Comparing with Fig.~\ref{fig:modes0}, these are precisely the modes with rotational symmetry, which perfectly matches the finding in collisions at fixed vanishing impact parameter in Ref.~\cite{Borghini:2022iym}.
The relative energy content of the other modes with $l\leq 39$ is smaller by about an order of magnitude --- for the modes $l=12$ and $l=25$ --- or more. 

In contrast, 12 of the 40 modes in mid-peripheral collisions have a sizable $|\tilde{\varepsilon}_0|_l$, ranging from 1 to 4\% of the energy content of the average initial state, but up to almost 12\% for the mode $l=0$ and close to 8\% for the modes $l=2$ and $l=5$. 
These three modes thus carry a significant energy content. 
This is a novel feature compared to the findings in collisions at fixed finite impact parameter in Ref.~\cite{Borghini:2022iym}: for events at $b=9$~fm, it was found that a number of modes carry a few percent of the energy content of the average initial state, as is the case here in the 30--40\% centrality class, but none of them had a value $|\tilde{\varepsilon}_0|_l$ larger than 4\%. 
Accordingly, it seems that the fluctuation modes $l=0$, $l=2$ and $l=5$ for events in the 30--40\% centrality class account (partly) for a type of fluctuation that is absent in collisions at fixed impact parameter:\footnote{This impression is reinforced by the fact that in Ref.~\cite{Borghini:2022iym} no mode was found with a profile similar to those of the modes $l=2$ and $l=5$ (as well as $l=8$ and 14) of Fig.~\ref{fig:modes30}: also visually these modes look new compared to those at fixed $b=9$~fm.} 
the natural guess is that they are to some extent reflecting the impact-parameter variation across the events of the centrality class under consideration. 
At the same time, these modes are also affected by the fluctuations in the positions of the nucleons inside the colliding nuclei.
While it is clear that the two sources of fluctuations present in the MC Glauber model cannot be disentangled, we performed a short investigation of fluctuation modes in an optical Glauber model, in which only the geometry fluctuations from the varying impact parameter occur. 
We present the corresponding results in Appendix~\ref{apx:optical-Glauber}, together with our speculation of how the salient features of the optical-Glauber modes could survive in the modes of the MC Glauber model.

\begin{figure*}
	\includegraphics[width=.495\linewidth]{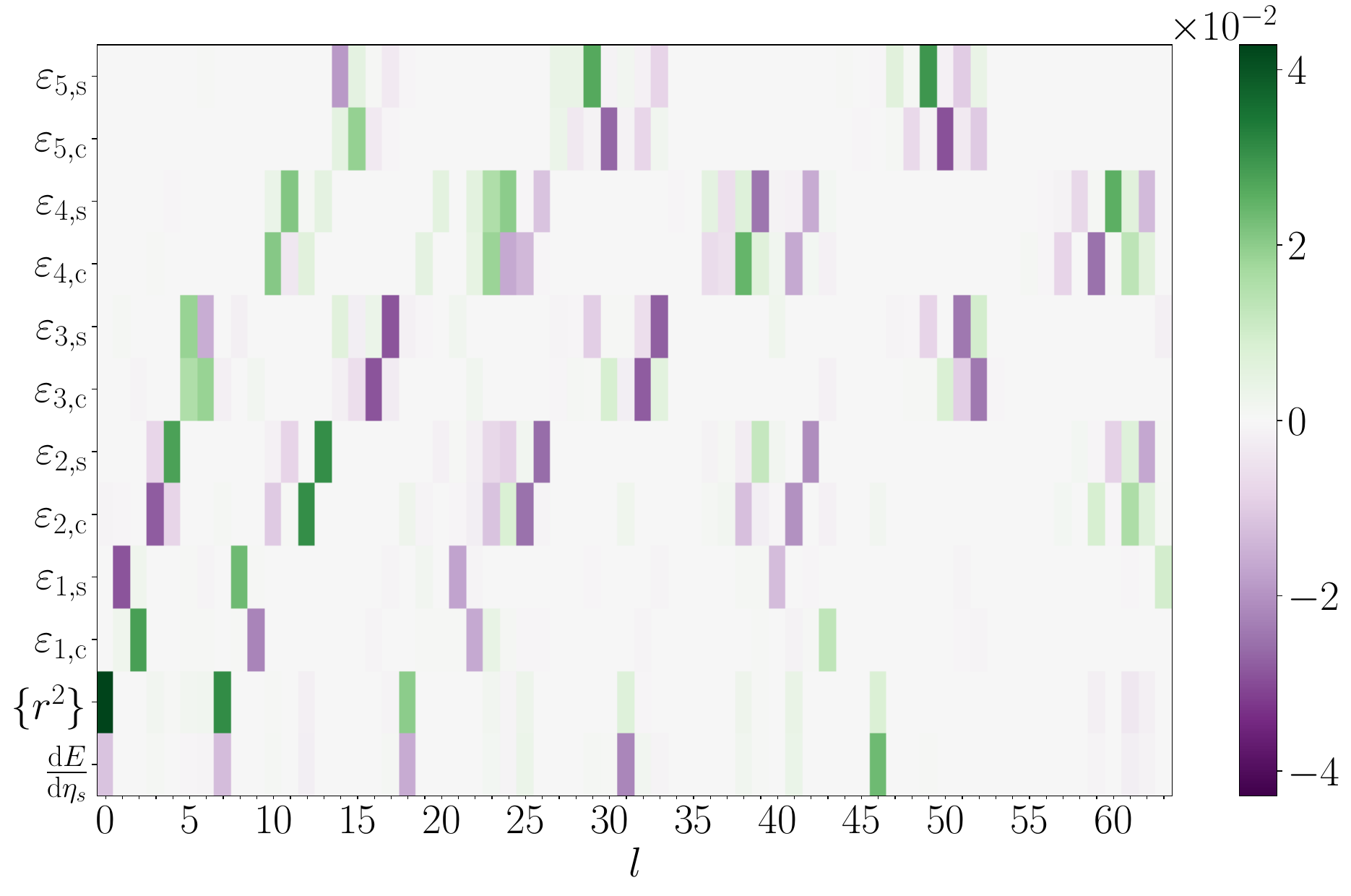}\hfill
	\includegraphics[width=.495\linewidth]{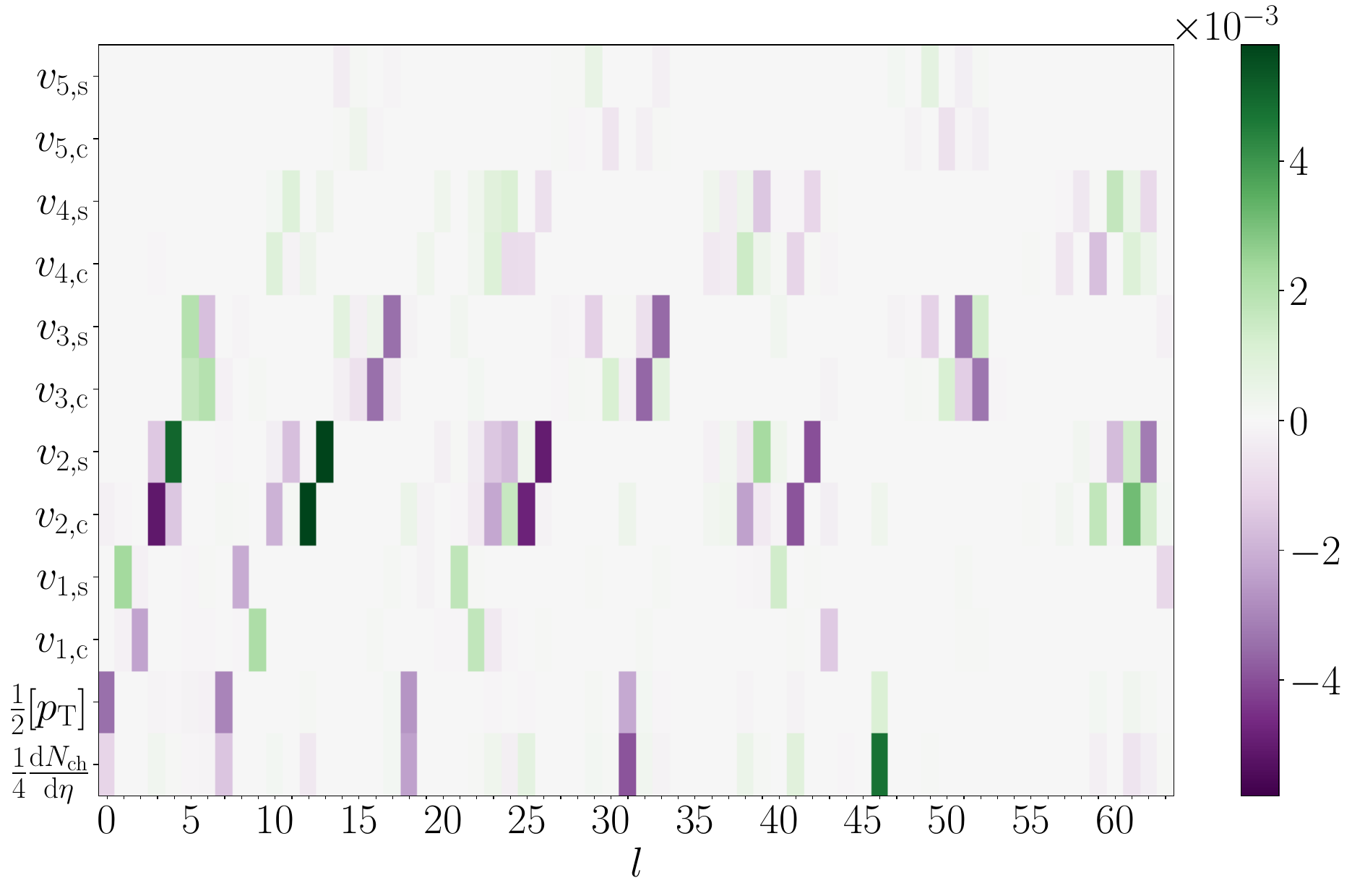}\vspace{-3mm}%
	\caption{Linear-response coefficients $L_{\alpha,l}$ for initial-state characteristics (left) and final-state observables computed from the \MUSIC\ spectra (right) for the first 64 modes for events in the 0--2.5\% centrality class. The coefficients for dimensionful observables and for multiplicity have been divided by $\bar{O}_\alpha$.}
	\label{fig:L_IS+FS_0}
\end{figure*}

\subsection{Mode-by-mode evolution: Initial-state observables}
\label{sec:IS}

In Sec.~\ref{ss:response_theory} we introduced the formalism underlying the mode-by-mode analysis of observables, based on expansion~\eqref{eq:observables} with the coefficients~\eqref{eq:def_L_l} and \eqref{eq:def_Q_ll'} characterizing the linear and quadratic response. 
In the present and the following section we apply this idea to a number of observables in the initial and in the final state of a collision. 

It is important to note that the formalism does not apply to any observable, but only to those smooth enough to ensure that expansion~\eqref{eq:observables} holds. 
In particular, the usual spatial eccentricities $\varepsilon_n$ defined (together with the symmetry-plane angle $\Phi_n$) by~\cite{Teaney:2010vd,Gardim:2011xv}%
\begin{equation}
\varepsilon_1 \ee^{\ii\Phi_1} \equiv 
	-\frac{\displaystyle\int\! r^3 \ee^{\ii\theta}e(r,\theta)\,r\,\dd r\,\dd\theta}%
	{\displaystyle\int\! r^3 e(r,\theta)\,r\,\dd r\,\dd\theta}
	\quad\text{for } n=1 \label{eq:epsilon_1}
\end{equation}
and
\begin{equation}
\varepsilon_n \ee^{\ii n\Phi_n} \equiv 
  -\frac{\displaystyle\int\! r^n \ee^{\ii n\theta}e(r,\theta)\,r\,\dd r\,\dd\theta}%
	{\displaystyle\int\! r^n e(r,\theta)\,r\,\dd r\,\dd\theta}
	\quad\text{for } n\geq 2,
	\label{eq:epsilon_n}
\end{equation}
do {\it not\/} obey Eq.~\eqref{eq:observables} when the value of $\varepsilon_n$ of the average initial state $\bar{\Psi}$ is zero --- which is the case for all odd eccentricities. 
In contrast, the real and imaginary parts of $\varepsilon_n \ee^{\ii n \Phi_n} = \varepsilon_{n, c} + \ii_{}\varepsilon_{n, s}$ do satisfy Eq.~\eqref{eq:observables} and will be among the observables we consider in the initial state, for $n\leq 5$.

Another set of observables that obey Eq.~\eqref{eq:observables} are the moments in $r$ of $e(\bm{x})$, i.e.\ the energy-density-weighted averages $\{r^k\}$ for any $k\geq 1$. 
By construction $\{r\}=0$, since all our initial states are re-centered and $r$ is defined from this common center. 
Hereafter, we only present the mean square radius $\{r^2\}$.

Eventually, we also consider the energy per unit space-time rapidity, $\dd E/\dd\eta_s$, which is given by the product by the initial time $\tau_0$ of the integral of $e(\bm{x})$ over the transverse plane. 
Note that since $\dd E/\dd\eta_s$ is strictly linear --- the energy of $\bar{\Psi} + \xi\Psi_l$ equals the sum of the energy of $\bar{\Psi}$ and of $\xi$ times the energy of $\Psi_l$ ---, the corresponding linear-response coefficient $L_{\dd E/\dd\eta_s,l}$ is precisely the energy of $\Psi_l$. 
Dividing this coefficient by the energy of $\bar{\Psi}$ thus yields the relative energy content, whose modulus $|\tilde{\varepsilon}_0|_l$ we discussed in the previous section. 

Technically, we use the discretized derivatives~\eqref{eq:L&Q_num} to compute the linear and quadratic response coefficients, where $O^\pm_{\alpha, l} \equiv O_\alpha(\Bar{\Psi} \pm \xi \Psi_l)$.
For observables that are strictly linear, any value of the parameter $\xi$ may be used, but this is no longer true if $O_\alpha$ is not linear, as is the case of most observables we consider. 
Accordingly, one has to find a convenient value of $\xi$: if it is too large, then the nonlinear terms start to be important. 
But if $\xi$ is too small, the difference between $O^+_{\alpha, l}$ and $O^-_{\alpha, l}$ may be dominated by numerical noise. 
Testing several values of $\xi$, we checked empirically that any choice in the range $[0.01,0.5]$ yields reliable derivatives for the observables we consider both in the initial and the final state. 
While in Ref.~\cite{Borghini:2022iym} we used 0.1, here the response coefficients we report were computed with $\xi=0.5$, for a reason that will be detailed in Sec.~\ref{sec:FS_SMASH}.

\begin{figure*}
    \includegraphics[width=.495\linewidth]{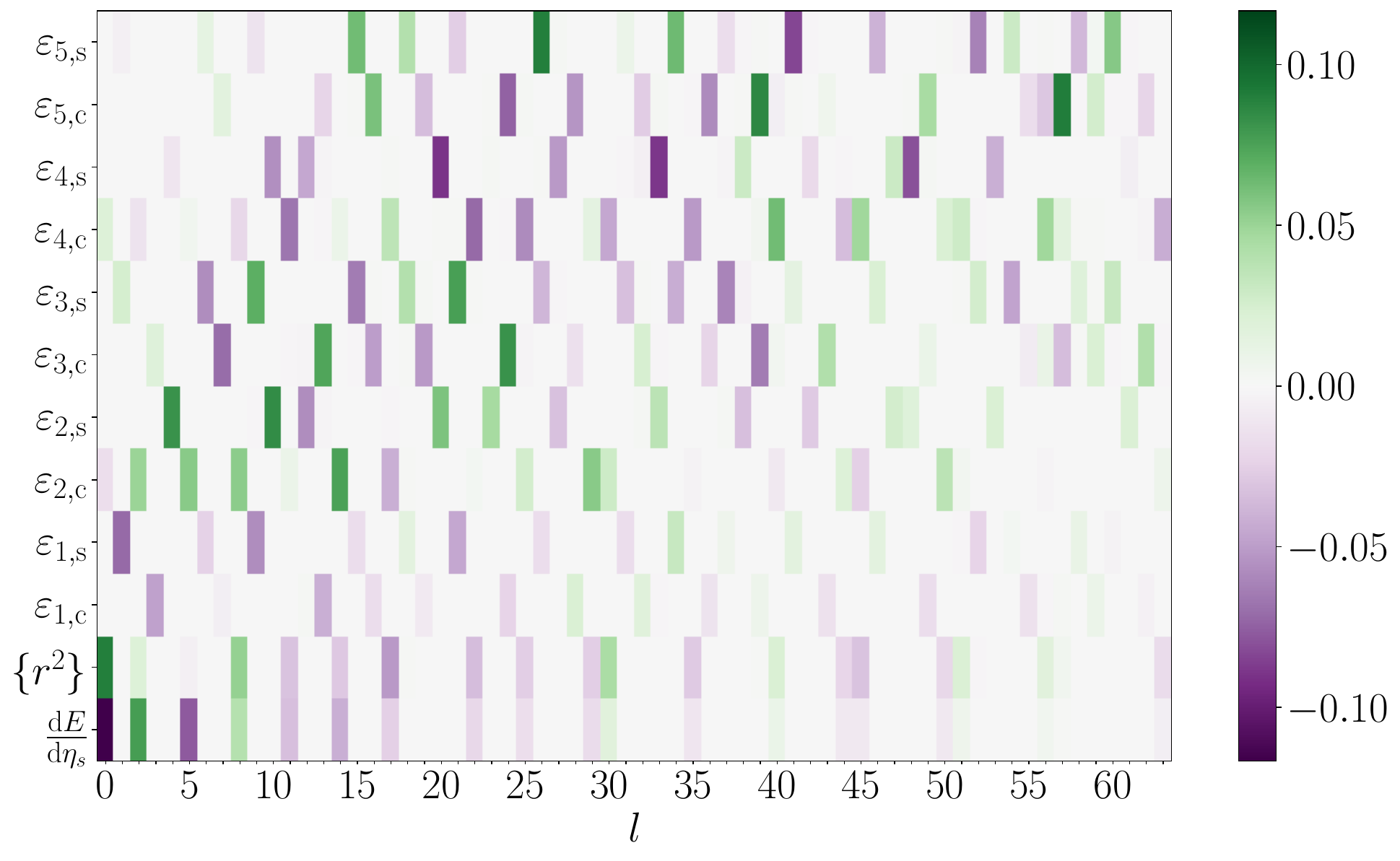}\hfill
	\includegraphics[width=.495\linewidth]{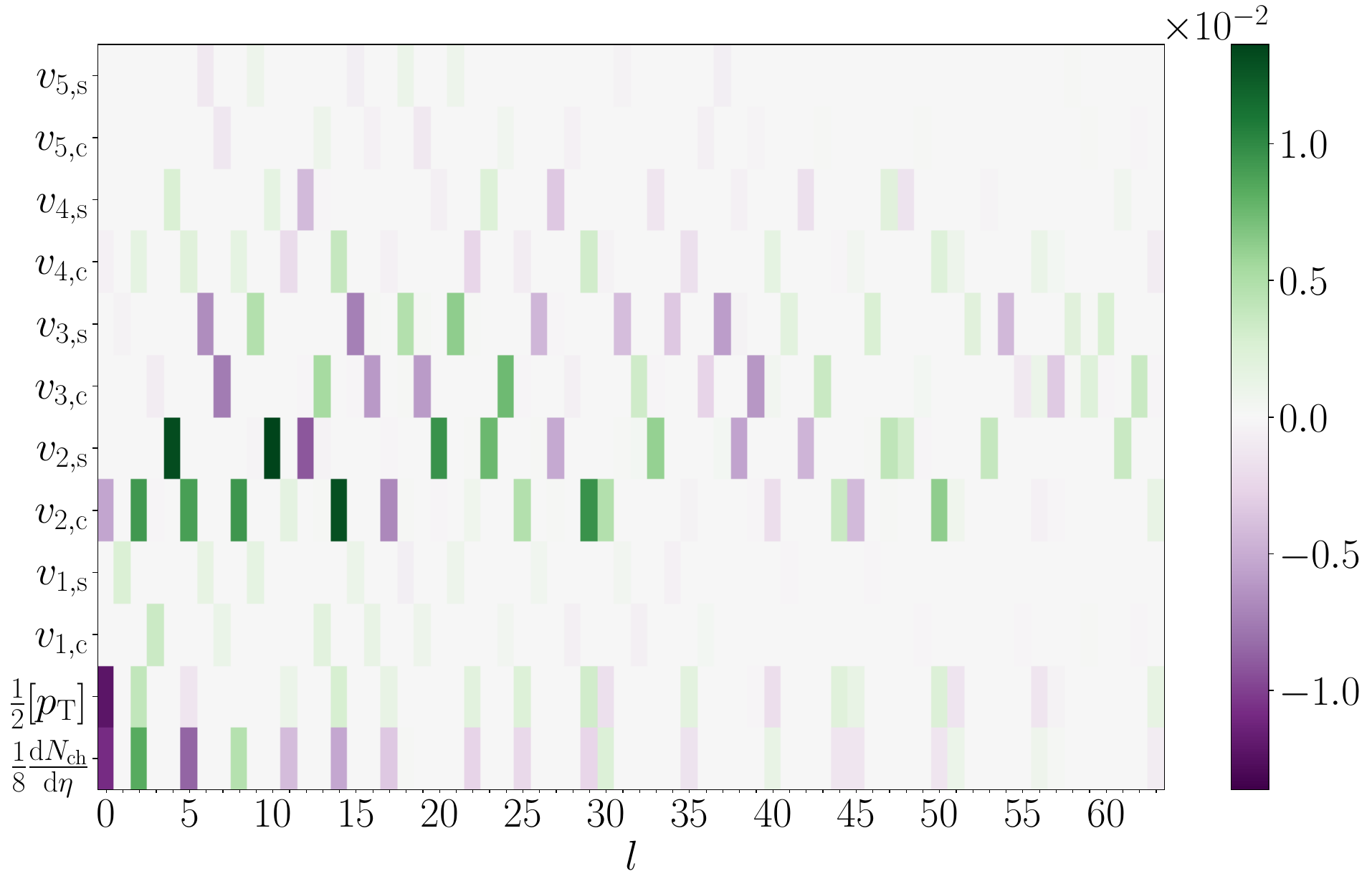}\vspace{-3mm}%
	\caption{Linear-response coefficients $L_{\alpha,l}$ for initial-state characteristics (left) and final-state observables computed from the \MUSIC\ spectra (right) for the first 64 modes for events in the 30--40\% centrality class. The coefficients for dimensionful observables and for multiplicity have been divided by $\bar{O}_\alpha$.}
	\label{fig:L_IS+FS_30}
\end{figure*}

In the left panel of Fig.~\ref{fig:L_IS+FS_0} we show the linear-response coefficients $L_{\alpha, l}$ for the initial-state characteristics we listed above, computed for the first 64 fluctuation modes for central events. 
In turn, the left panel of Fig.~\ref{fig:L_IS+FS_30} shows the same coefficients for events in the 30--40\% centrality class. 
To obtain dimensionless numbers, the coefficients for $\dd E/\dd\eta_s$ and $\{r^2\}$ are divided by the respective value of the observable in the average state. 

Starting with the central events, one first finds again that only few of the modes contribute to the energy per unit space-time rapidity, as already discussed in Sec.~\ref{sec:mode-results}. 
Here, we see that these modes with energy content are the only ones that change the mean square radius of the system, and also that they mostly have vanishing eccentricities $\varepsilon_{n,\rm c/s}$, as could be expected from their circular symmetry we recognized in Fig.~\ref{fig:modes0}. 
There are exceptions to this ``rule'', e.g.\ the modes with $l=12$, 25, 38, 42\dots, that actually have very little energy content and a shape that by eye is not rotationally symmetric. 

Regarding the eccentricities, the situation differs from the modes at fixed vanishing impact parameter in Ref.~\cite{Borghini:2022iym}: 
At $b=0$ the modes (generally) only have an eccentricity in a single harmonic, but with both $\varepsilon_{n,\rm c}$ and $\varepsilon_{n,\rm s}$, since the $x$-axis plays no special role. 
Here in contrast the fluctuation modes tend to have eccentricities in different harmonics — of a given parity —, and also they more often have only the cosine or the sine part, or at least there is a clear hierarchy between $\varepsilon_{n,\rm c}$ and $\varepsilon_{n,\rm s}$. 
Both features are caused by the variation in impact parameter in the centrality class, even if it is small in the central bin. 
Similar to the finding in collisions at $b=0$, one finds that non-circular modes tend to come in pairs, one carrying $\varepsilon_{n,\rm c}$ and the other $\varepsilon_{n,\rm s}$, with the same magnitude in absolute value.

Turning to the modes for collisions in the 30--40\% centrality bin, one first recognizes that the modes that contribute to $\dd E/\dd\eta_s$ are more numerous than in central collisions, as already seen in Sec.~\ref{sec:mode-results}, but again they are the only modes that contribute to $\{r^2\}$.
In addition, they also contribute to $\varepsilon_{2,\rm c}$ and $\varepsilon_{4,\rm c}$, but not to any odd $\varepsilon_{n,\rm c}$ nor to the sine parts $\varepsilon_{n,\rm s}$. 
This matches the properties of the average initial state $\bar{\Psi}$ itself, which has non-zero $\varepsilon_{2,\rm c}$ and $\varepsilon_{4,\rm c}$, and vanishing other eccentricities (at least, with $n\leq 5$).

\begin{figure}[b]
	\includegraphics[width=\linewidth]{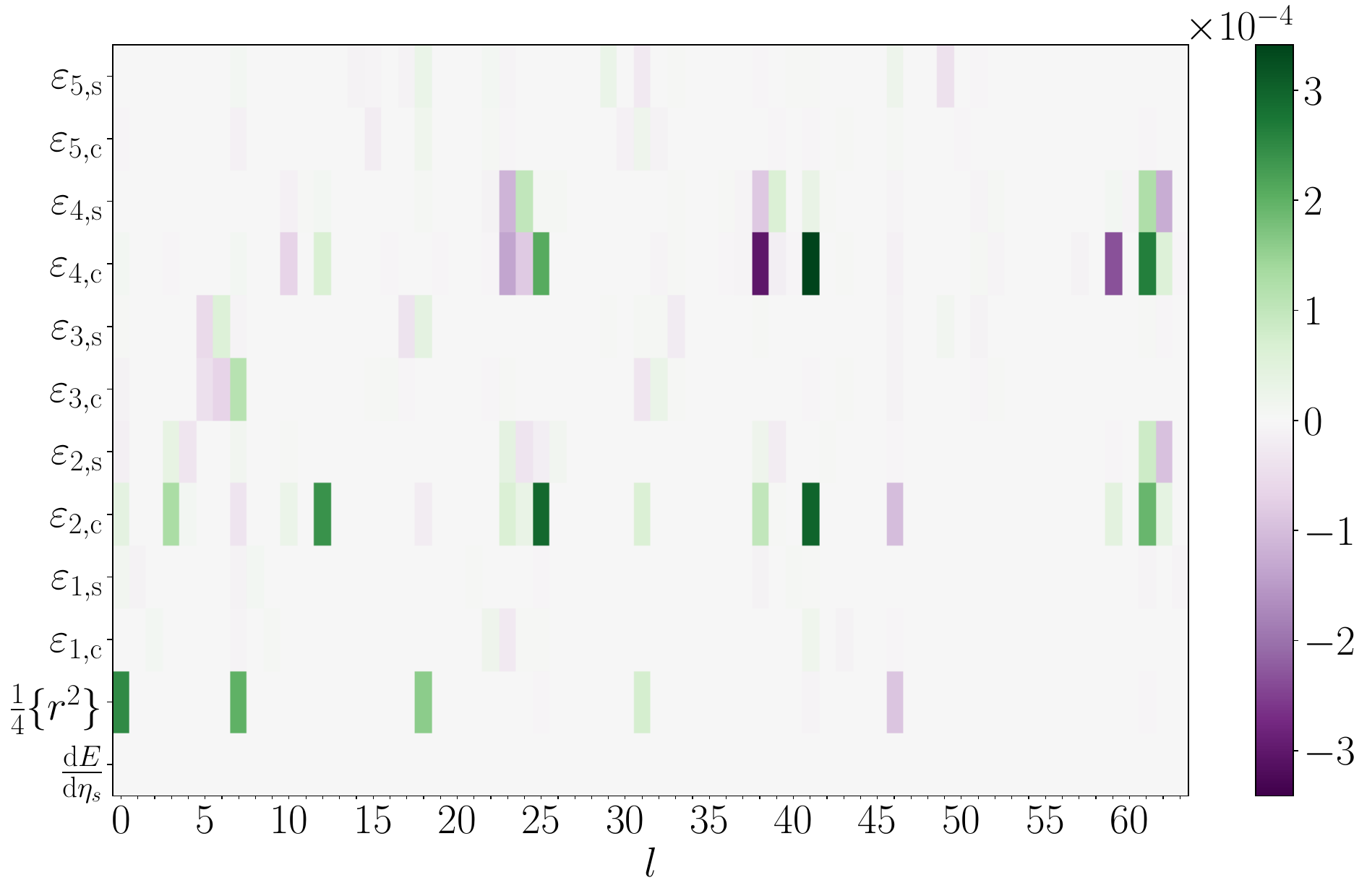}
	\includegraphics[width=\linewidth]{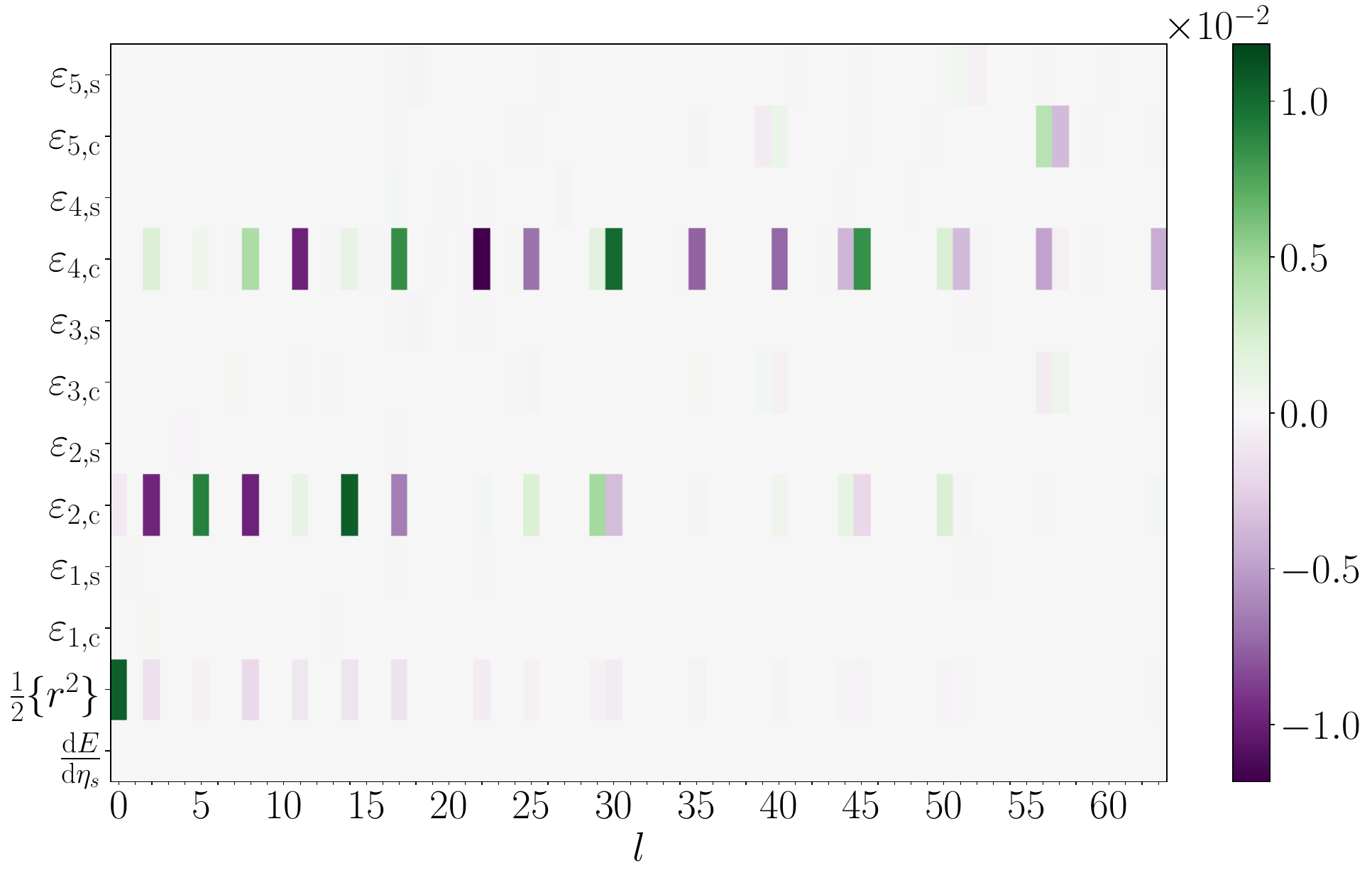}\vspace{-3mm}
	\caption{Quadratic-response coefficients $Q_{\alpha,ll}$ for initial-state characteristics for the first 64 modes for events in the 0--2.5\% (top)  and 30--40\% (bottom) centrality class. 
		The coefficients for dimensionful observables have been divided by $\bar{O}_\alpha$.}
	\label{fig:Q_IS}
\end{figure}

The other modes, without any energy content, tend generally to have all eccentricities of a given parity and with a given $\theta\to -\theta$ symmetry, other than non-zero $\varepsilon_{2,\rm c}$ and $\varepsilon_{4,\rm c}$.
Thus, there are modes with non-zero $\varepsilon_{2,\rm s}$ and $\varepsilon_{4,\rm s}$ (e.g.\ $\Psi_{l=4}$), modes like $\Psi_{l=1}$ with non-zero $\varepsilon_{1,\rm s}$, $\varepsilon_{3,\rm s}$, and $\varepsilon_{5,\rm s}$, or modes like $\Psi_{l=3}$ with non-zero $\varepsilon_{1,\rm c}$, $\varepsilon_{3,\rm c}$, and $\varepsilon_{5,\rm c}$. 
Note also that the contribution to $\varepsilon_{n,\rm c/s}$ tends to become smaller with increasing mode number $l$, which is clearly recognizable for $n=1$ or 2. 
This latter feature, which also holds for the energy per unit space-time rapidity and the mean square radius, means that only a few modes --- among the first ones --- contribute significantly to a given observable, which is a desirable property of the mode decomposition~\cite{Borghini:2022iym}. 

In Fig.~\ref{fig:Q_IS} we display the quadratic-response coefficients $Q_{\alpha,ll}$ for our initial-state observables, with those for central events in the top panel and those for mid-peripheral initial states in the bottom panel. 
As in the case of the linear-response coefficients, the quadratic coefficients for $\dd E/\dd\eta_s$ and $\{r^2\}$ are divided by the respective value of the observable in the average state. 

The first notable feature is that the $\{Q_{\alpha,ll}\}$ are much smaller than the $\{L_{\alpha,l}\}$: by two orders of magnitude in central events, and by one order of magnitude in the 30--40\% centrality class. 
As we mentioned above, $\dd E/\dd\eta_s$ is a linear observable, so the corresponding quadratic-response coefficient should be exactly zero, which is what we find to the precision of our calculation. 
The modes that contribute to mean square radius at linear order also contribute at quadratic order. 
Among the eccentricities, only $\varepsilon_{2,\rm c}$ and $\varepsilon_{4,\rm c}$ (up to a few exceptions) show a sizable quadratic response.
This is consistent with the fact that here nonlinearities can only come from the denominator in the definitions~\eqref{eq:epsilon_1}--\eqref{eq:epsilon_n}, and are actually due to the non-zero $\varepsilon_{2,\rm c}$ and $\varepsilon_{4,\rm c}$ of the average initial state.

\subsection{Mode-by-mode evolution: Final-state observables}
\label{sec:FS}

A purpose of the decomposition of the initial states in uncorrelated fluctuation modes is to assess the impact of each mode on final-state observables at the end of a dynamical evolution. 
Matching the quantities we investigated in the initial state, we consider a few global final-state observables: 
the charged multiplicity per unit pseudorapidity $\dd N_{\rm ch}/\dd\eta$, the event-by-event average transverse momentum $[p_{\rm T}]$ of particles --- where square brackets denote an average with the charged-hadron momentum distribution $\dd N_{\rm ch}/p_{\rm T}\,\dd p_{\rm T}\,\dd\varphi_{\rm p}\,\dd\eta$, with $\varphi_{\rm p}$ the azimuthal angle of (transverse) momentum ---, and a number of anisotropic-flow coefficients $v_{n,\rm c}$, $v_{n,\rm s}$ with $n\leq 5$, which are respectively the real and imaginary parts of%
\begin{equation}
v_n \ee^{\ii n\psi_n} \equiv 
  \frac{\displaystyle\int\! \ee^{\ii n\varphi_{\rm p}} \frac{\dd N_{\rm ch}}{p_{\rm T}\,\dd p_{\rm T}\,\dd\varphi_{\rm p}\,\dd\eta}\, p_{\rm T}\,\dd p_{\rm T}\,\dd\varphi_{\rm p}}%
	{\displaystyle\int\! \frac{\dd N_{\rm ch}}{p_{\rm T}\,\dd p_{\rm T}\,\dd\varphi_{\rm p}\,\dd\eta}\, p_{\rm T}\,\dd p_{\rm T}\,\dd\varphi_{\rm p}}.
	\label{eq:flow_harmonics}
\end{equation}
The reason for investigating the $\{v_{n,\rm c}, v_{n,\rm s}\}$ instead of $v_n$ is the same as for the spatial eccentricities: the cosine and sine parts satisfy Eq.~\eqref{eq:decomposition}, and thus allow us to meaningfully define linear and quadratic response, while the modulus does not. 

As described in Sec.~\ref{sec:numerical}, we consider two setups, both starting with \Kompost\ and \MUSIC, but that differ after particlization: 
in the first scenario, which we discuss in Sec.~\ref{sec:FS_MUSIC}, we directly use the output of \MUSIC\ to compute observables. 
In the second setup, the final-state observables are calculated at the end of a further evolution stage, namely with SMASH (Sec.~\ref{sec:FS_SMASH}).

\subsubsection{Observables from \MUSIC\ outputs}
\label{sec:FS_MUSIC}

After determining the particlization hypersurface at $T_{\rm f.o.}=155$\,MeV, the \MUSIC\ code allows one to compute momentum distributions of emitted particles: 
using the built-in \MUSIC\ routines, we determined these distributions for 320 particle species, which are then allowed to decay, without further scattering. 
After these decays, we obtain the momentum distribution $\dd N_{\rm ch}/p_{\rm T}\,\dd p_{\rm T}\,\dd\varphi_{\rm p}\,\dd\eta$ from which we can determine observables. 
From there, we can compute the linear and quadratic-response coefficients for the various observables, which we now discuss. 

\begin{figure}
    \includegraphics[width=\linewidth]{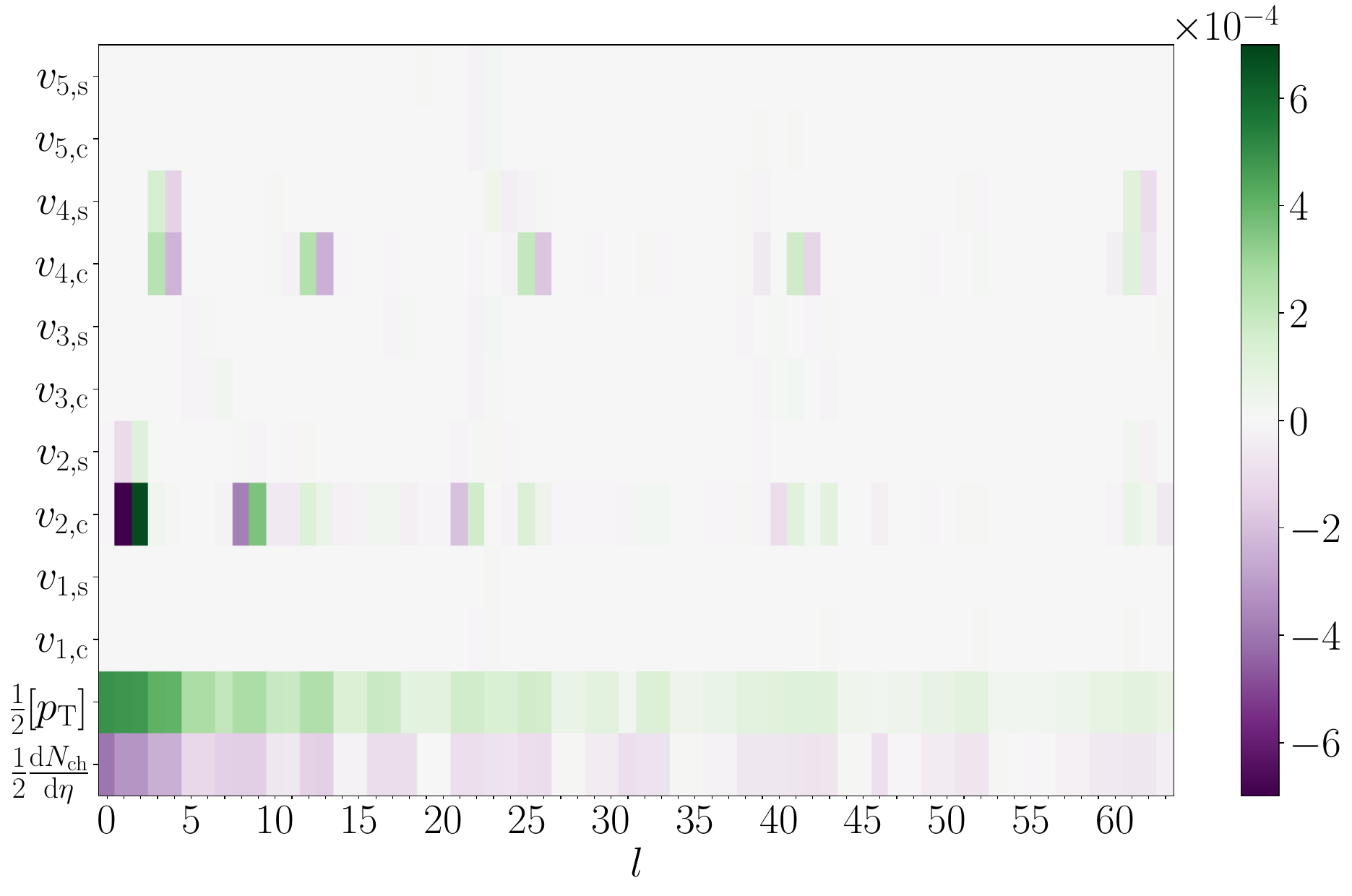}
    \includegraphics[width=\linewidth]{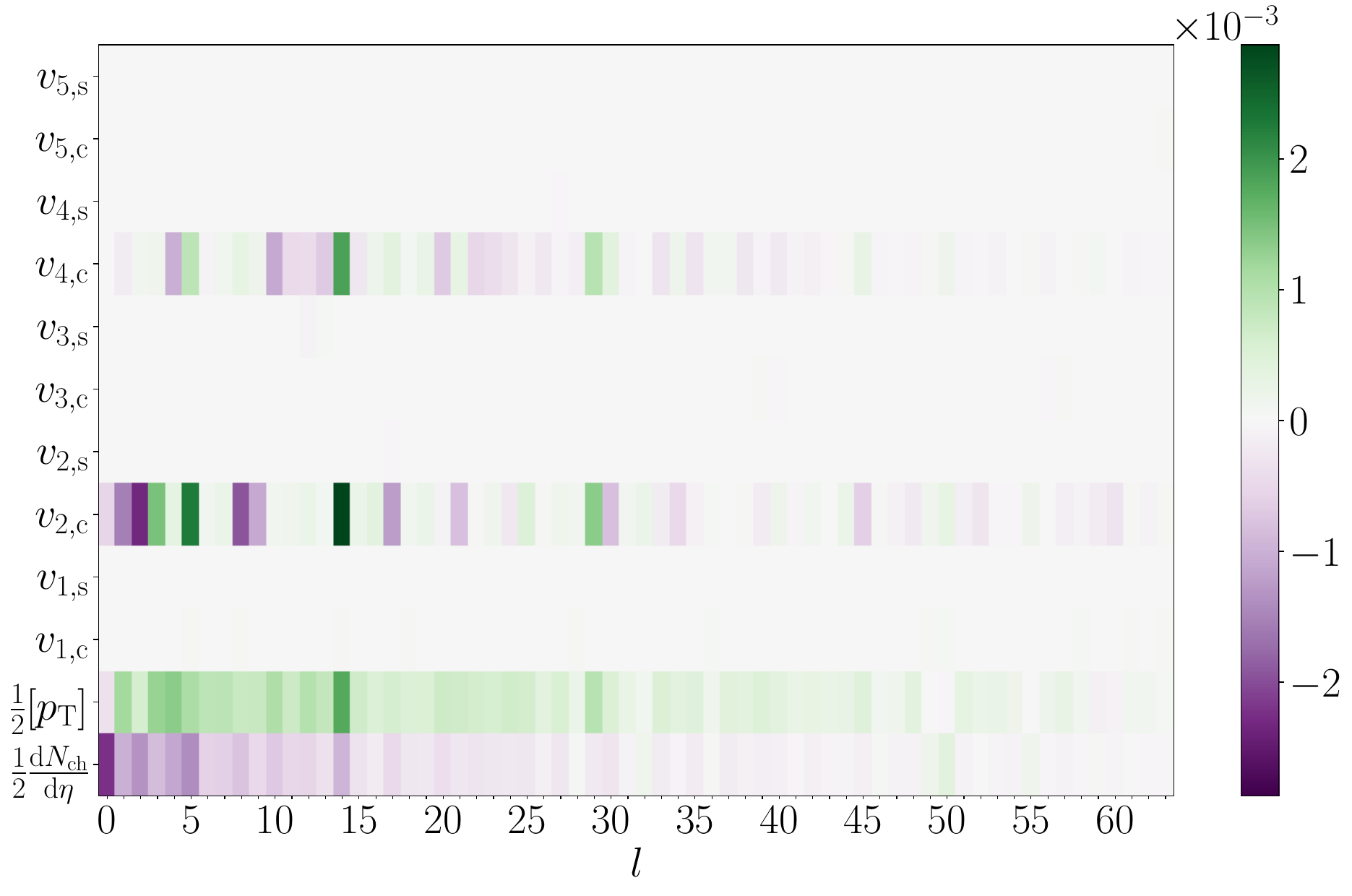}\vspace{-3mm}
    \caption{Quadratic-response coefficients $Q_{\alpha,ll}$ for final-state observables computed from the \MUSIC\ spectra for the first 64 modes for events in the 0--2.5\% (top) and 30--40\% (bottom) centrality class. 
    The coefficients for multiplicity and $[p_{\rm T}]$ have been divided by $\bar{O}_\alpha$.}
    \label{fig:Q_FS}
\end{figure}

The linear-response coefficients $L_{\alpha, l}$ for the final-state observables are displayed in the right panels of Fig.~\ref{fig:L_IS+FS_0} and Fig.~\ref{fig:L_IS+FS_30} for the first 64 fluctuation modes for events in the 0--2.5\% and 30--40\% centrality, respectively. 
In turn, Fig.~\ref{fig:Q_FS} shows the diagonal quadratic-response coefficients $Q_{\alpha,ll}$, with central events in the top panel and mid-peripheral ones in the bottom panel. 
To obtain dimensionless numbers of comparable magnitude, the coefficients for average transverse momentum are divided by twice the value of $[p_{\rm T}]$ in the average state, while the coefficients for $\dd N_{\rm ch}/\dd\eta$ are divided by the average-state multiplicity (multiplied by a number that can be read off the respective plot).

Comparing the left and right panels of Fig.~\ref{fig:L_IS+FS_0} or Fig.~\ref{fig:L_IS+FS_30}, one sees that at the linear level there is a one-to-one contribution between some initial-state characteristic and a final-state observable. 
Thus the modes that contribute linearly to $\dd E/\dd\eta_s$ also contribute linearly to multiplicity, while no other mode contributes. 
The same holds for $\{r^2\}$ and $[p_{\rm T}]$, and for every $\varepsilon_{n,\rm c/s}$ and the corresponding $v_{n,\rm c/s}$. 
There seems to be the general trend that the linear-response coefficient for $\dd E/\dd\eta_s$ and $\dd N_{\rm ch}/\dd\eta$ for a given mode have the same sign, as already observed in Ref.~\cite{Borghini:2022iym}.
This also holds for the coefficients of a given eccentricity and the associated flow coefficient for $n\geq 2$. 
In contrast, the linear coefficients for $\varepsilon_{1,\rm c/s}$ and the associated $v_{1,\rm c/s}$ have opposite signs. 
On the other hand, there does not seem to be a constant relationship between the signs of the coefficients for mean square radius and average transverse momentum, at least in the 30--40\% centrality class. 
In central collisions, there seems to be an anticorrelation between these coefficients, which can readily be understood: 
if a mode tends to decrease the mean square radius ($L_{\{r^2\},l}<0$), then since the initial state will be ``denser'' (assuming the total energy does not change by too much), leading to greater pressure gradients, that will lead to a greater acceleration of the expanding system, and thus to an increased average momentum $L_{[p_{\rm T}],l}>0$ in the final state.
This reasonable expectation for central events with their almost fixed size may then be blurred in the 30--40\% centrality bin, in which events have a different geometry, leading to the loss of anticorrelation between $\{r^2\}$ and $[p_{\rm T}]$.

An easily understandable feature is the trend that the linear response of the anisotropic-flow harmonics becomes weaker with growing $n$, reflecting the increasing influence of viscous damping. 
There is also a similar pattern with increasing mode number $l$. 

Regarding the quadratic-response coefficients, the trends are rather clear and match the findings in Ref.~\cite{Borghini:2022iym}. 
All modes contribute quadratically to multiplicity and average transverse momentum. 
In central collisions, all modes with a dipole asymmetry $\varepsilon_1$ contribute to $v_{2,c}$ and all those with an ``ellipticity'' $\varepsilon_2$ contribute to $v_{4,c}$: this reflects the known quadratic contribution of $\varepsilon_n$ to $v_{2n}$ in hydrodynamics~\cite{Borghini:2005kd,Gardim:2011xv,Teaney:2012ke,Niemi:2012aj}.
In the 30--40\% centrality class, basically all modes have non-vanishing quadratic coefficients for $v_{2,c}$, $v_{4,c}$.
Most of the remaining quadratic coefficients are almost vanishing.

\subsubsection{Observables from SMASH outputs}
\label{sec:FS_SMASH}

Instead of letting only the hadron decays, a more realistic and state-of-the-art approach is to allow them to rescatter, using an afterburner like SMASH~\cite{SMASH:2016zqf}, which may modify the values of the final observables. 
While this is physically well motivated, this approach also comes with a drawback. 
In contrast to the momentum distributions produced by \MUSIC, which for practical purposes amount to having a very large number of particles per event, the multiplicity of charged hadrons in SMASH --- which we shall momentarily denote with $M$ --- is the realistic one: say, about 2000 (per unit pseudo-rapidity) in central events and 500 for events in the 30--40\% centrality class. 
This entails a typical statistical ``noise'' of order $1/\sqrt{M}$ on any observable. 
This can be problematic for the calculation of the response coefficients~\eqref{eq:L&Q_num}, as we shall illustrate on an example. 

From the right panel of Fig.~\ref{fig:L_IS+FS_30}, we see that the typical ``signal'' for a linear-response coefficient $L_{v,l}$ for a flow harmonic is of order $10^{-3}$--$10^{-2}$ in mid-peripheral collisions --- and a factor 3 smaller in central events, see Fig.~\ref{fig:L_IS+FS_0}. 
Given the numerical definition~\eqref{eq:L&Q_num} of $L_{v,l}$, this means that the values of $v_n(\bar{\Psi}+\xi\Psi_l)$ and $v_n(\bar{\Psi}-\xi\Psi_l)$ that we compute differ by $2\xi$ times this typical magnitude. 
That sets the maximal acceptable ``noise'' on the computation of the individual $v_n(\bar{\Psi}\pm\xi\Psi_l)$, which should be about an order of magnitude smaller: to fix ideas, to be able to compute reliably a coefficient $L_{v,l} \approx 10^{-3}$, the numerical noise should be (at most) of order $10^{-4}$ times the value of $\xi$. 

With the experimental value of multiplicity $M$, this clearly does not hold. 
Yet in hybrid approaches it is usual to perform oversamplings at the particlization hypersurface, producing $N_{\rm samples}$ ``particle events'' from a single hydrodynamical event. 
Accordingly, the effective multiplicity from which the numerical noise should be estimated is $N_{\rm samples}\cdot M$.
Here we want a numerical noise of order $10^{-4}$ --- with a value $\xi=0.5$ ---, $N_{\rm samples}\cdot M$ must be of order $10^8$: with $M\approx 500$ particles in mid-peripheral events, there has to be (at least) $N_{\rm samples}=20,000$ oversamplings, which is the choice we made. 
Note that with the five times smaller value $\xi=0.1$ that was used in Ref.~\cite{Borghini:2022iym}, one should produce five times as many oversamplings, which is why we chose $\xi=0.5$.

\begin{figure}
	\includegraphics[width=\linewidth]{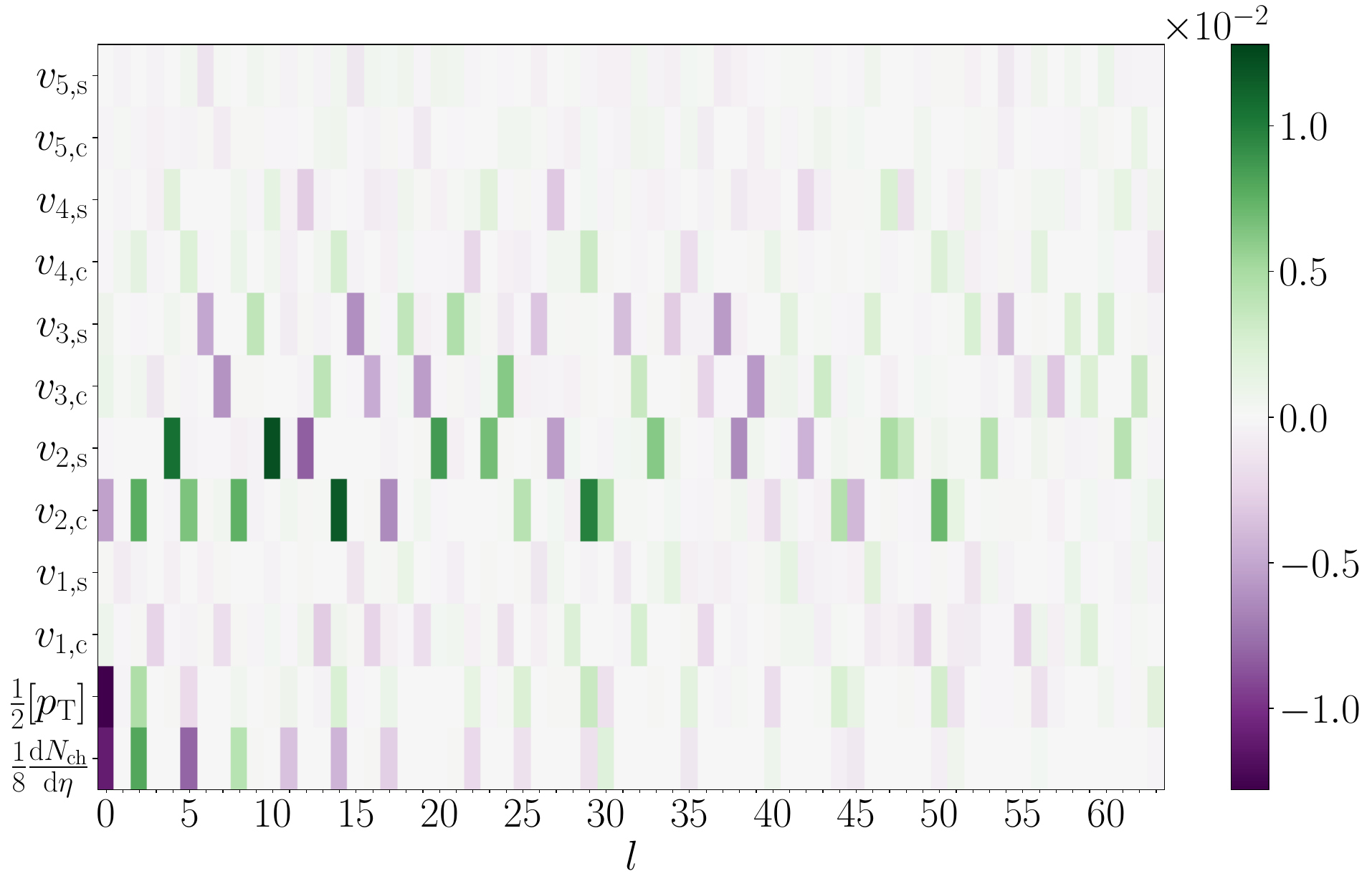}\vspace{-3mm}
	\caption{Linear-response coefficients $L_{\alpha,l}$ for final-state observables computed from the SMASH output for the first 64 modes for events in the 30--40\% centrality class. The coefficients for multiplicity and $[p_{\rm T}]$ have been divided by $\bar{O}_\alpha$.}
	\label{fig:L_FS_SMASH}
\end{figure}

Knowing that a significant number of oversamplings has to be performed to reliably compute the response coefficients necessary for mode-by-mode evolution, we used iSS~\cite{Shen:2014vra} and SMASH to produce hadrons and let them rescatter, and the SPARKX package~\cite{Sass:2025opk} to produce the final observables. 
Viewing this as a proof-of-principle investigation --- to our knowledge, no similar study of the propagation of ``modes'' of initial-state fluctuation through the state-of-the-art pipeline pre-equilibrium + hydrodynamic evolution + hadronic afterburner has ever been reported ---, we only present results for events in the 30--40\% centrality class.\footnote{In central collisions the larger event multiplicity helps reducing the statistical noise, but on the other hand the target coefficients are smaller, so that one needs about the same number of oversamplings to obtain the same precision. Since SMASH events with larger multiplicity require more running time, we chose to test only the mid-peripheral collisions.}

Figure~\ref{fig:L_FS_SMASH} shows the linear-response coefficients for modes in that centrality bin, for final-state observables computed from the SMASH output with 20,000 oversamplings. 
Comparing with the coefficients computed from the \MUSIC\ spectra, shown in the right panel of Fig.~\ref{fig:L_IS+FS_30}, we see that the results are quite similar. 
This means on the one hand that the numerical noise of the SMASH events is under control,\footnote{Note that we actually chose the overall scales of Figs.~\ref{fig:L_IS+FS_30} and \ref{fig:L_FS_SMASH} such that one starts to see the numerical noise in the latter --- mostly on $v_1$, $v_4$ and $v_5$.} and on the other hand that the considered observables are not strongly modified by the hadronic rescatterings. 
The only exception, which remains unexplained, is that the directed-flow coefficients $v_{1,\rm c/s}$ change sign between the \MUSIC\ and SMASH calculations.
Looking more carefully, we find differences at the per-mille level. 
For instance, $L_{v_{2,\rm c},l=5} = 8.9\times 10^{-3}$ for $v_{2,\rm c}$ in mode $l=5$ with the \MUSIC\ output, while the value from the SMASH output is $L_{v_{2,\rm c},l=5} = 6.4\times 10^{-3}$. 
There seems to be the generic trend that the coefficients for the flow harmonics $v_2$ and $v_3$ are smaller in absolute value at the end of the SMASH evolution.

\begin{figure}
	\includegraphics[width=\linewidth]{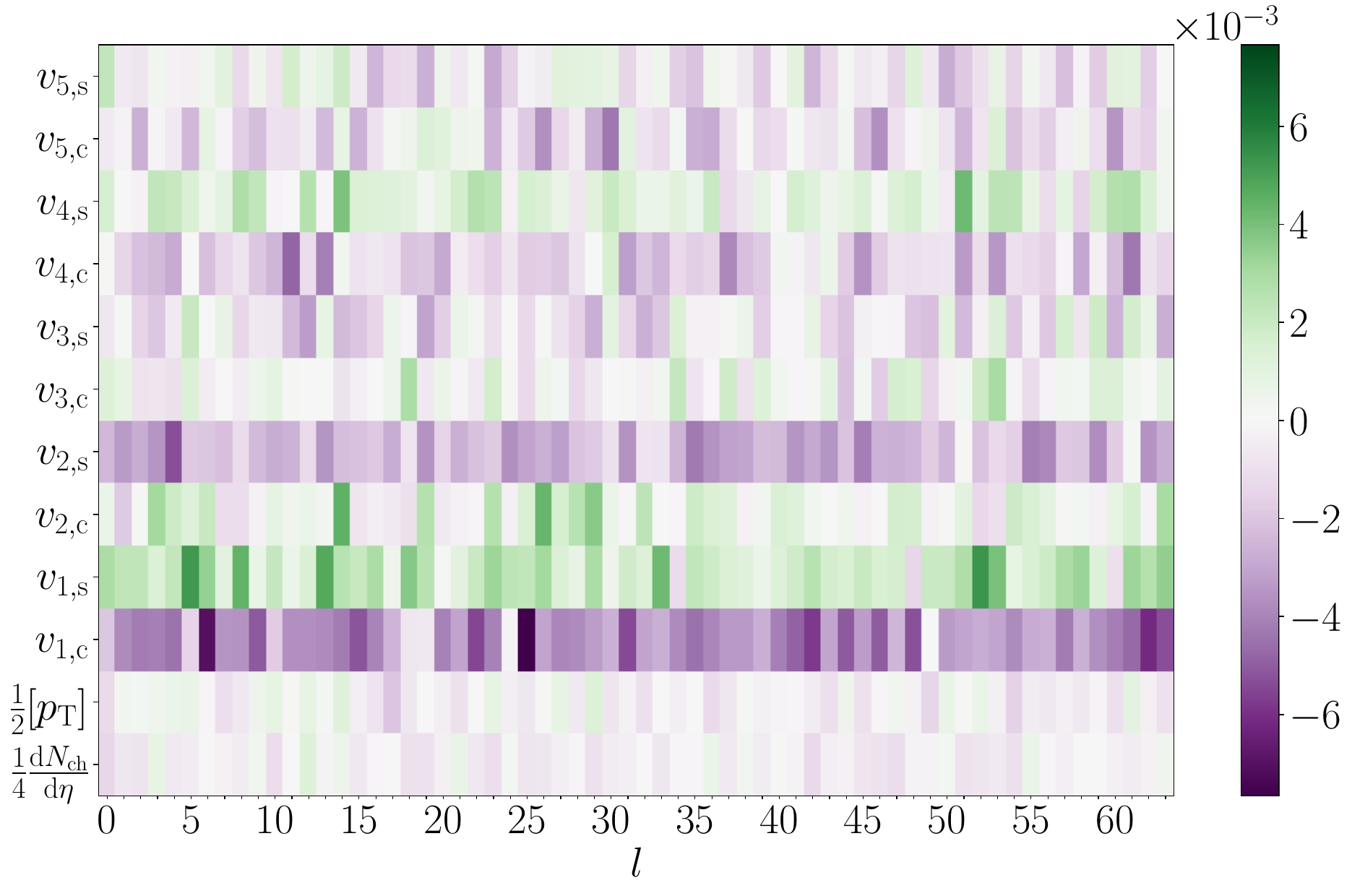}\vspace{-3mm}
	\caption{Quadratic-response coefficients $Q_{\alpha, ll}$ for final-state observables computed from the SMASH output for the first 64 modes for events in the 30--40\% centrality class. As discussed in the text, these coefficients are dominated by numerical noise.}
	\label{fig:Q_FS_SMASH}
\end{figure}

To illustrate the effect of numerical noise on the mode-by-mode analysis, we display in Fig.~\ref{fig:Q_FS_SMASH} the quadratic-response coefficients for final-state observables computed from the SMASH output, for the same modes as in Fig.~\ref{fig:L_FS_SMASH}.
From the bottom panel of Fig.~\ref{fig:Q_FS}, which shows the same coefficients computed from the \MUSIC\ output, we know that the typical magnitude of the $Q_{\alpha, ll}$ is about $10^{-4}$--$10^{-3}$, i.e.\ one order of magnitude smaller than the linear coefficients. 
But a few $10^{-4}$ is precisely the scale we set for the numerical noise on the $L_{\alpha,l}$,\footnote{Because of the different factors in the denominators of the numerical derivatives in Eq.~\eqref{eq:L&Q_num}, one already needs more precision on the observables to extract the second derivative than the first.} so one can anticipate that the calculated quadratic-response coefficients are plagued by noise, which is clearly what we find in Fig.~\ref{fig:Q_FS_SMASH}.
A rule-of-thumb estimate similar to that done above for the noise on the linear-response coefficients shows that one would need of the order of $10^7$ oversamplings to reduce the noise to an acceptable level here. 
Yet one should also note that the quadratic coefficients for observables that have a sizable signal in the average initial state --- like multiplicity, average transverse momentum, $v_{2,\rm c}$ or $v_{3,\rm c}$ --- are much less noisy than the coefficients for $v_{1,\rm c}$ or $v_{2,\rm s}$. 
Thus, if one wants to focus on such observables, the increase in the number of required oversampling may remain at an acceptable level.

\section{Discussion}
\label{sec:conclusions}

Using a nucleon-based MC Glauber model, we have determined the average state and the uncorrelated modes that generate the event-by-event fluctuations of the initial state in two centrality classes of Pb-Pb collisions at 5.02~TeV. 
We then investigated how each fluctuation mode affects the final state of the collisions, testing the feasibility of a state-of-the-art hybrid evolution with pre-equilibrium (\Kompost), dissipative fluid dynamics (\MUSIC), and a hadronic afterburner (SMASH). 
 
The fluctuation modes in mid-peripheral event (30--40\% centrality class) present a number of  differences with those at a fixed finite value of impact parameter studied in Ref.~\cite{Borghini:2022iym}. 
In particular, some of the most important modes carry a significant energy content, which according to us is a signal that they partly account for the variation in impact-parameter value across the centrality bin.
These modes also yield the largest contributions to the fluctuations of multiplicity and event-averaged transverse momentum in the final state.  
 
In contrast, the fluctuation modes in central events (0--2.5\%) are very similar to those previously determined at fixed vanishing impact parameter~\cite{Borghini:2022iym}. 
This nicely confirms the repeated claim that ultracentral events are to a very large extent free from the extra source of initial-state ``fluctuation'' due to impact-parameter variation, and thus represent a cleaner setup to probe physics of the created medium~\cite{Luzum:2012wu,Gardim:2019brr}.

A significant advantage of using uncorrelated fluctuation modes to expand the initial state, instead of a priori fixed modes, is the ability to push the study of the response to quadratic order at moderate cost. 
Here we demonstrated that it is also possible to investigate the response at the end of a full hybrid simulation including not only pre-equilibrium and hydrodynamical stages, but also a hadronic transport cascade as final stage. 
The main issue is the need to produce enough statistics to decrease the noise in the output for the evolution of the single-mode initial states $\Bar{\Psi} \pm \xi \Psi_l$ that underlie the mode-by-mode analysis. 
We showed that this is feasible, at the cost of a large number $N_{\rm samples}$ of oversamplings at the particlization hypersurface. 
A possible way to reduce $N_{\rm samples}$ is to use test particles in the afterburner, although this naturally increases the running time of each hadron-transport simulation. 
In the future, we plan to investigate more systematically how to optimize the simulation time, especially with a view to studying systems in which the assumption of longitudinal boost invariance is relaxed. 
This will be important to assess which features of genuinely 3D initial states are reflected in final-state observables.

\begin{acknowledgments}
	We would like to thank Benedikt Bachmann, Giuliano Giacalone, Aleksas Mazeliauskas, S\"oren Schlichting and Chun Shen for valuable discussions.
    The authors acknowledge support by the Deutsche Forschungsgemeinschaft (DFG, German Research Foundation) through the CRC-TR 211 'Strong-interaction matter under extreme conditions' - project number 315477589 - TRR 211.
    H.~R. was supported by the National Science Foundation (NSF) within the framework of the JETSCAPE collaboration (OAC-2004571) and by the DOE (DE-SC0024232).
Numerical simulations presented in this work were performed at the Paderborn Center for Parallel Computing (PC$^2$) and we gratefully acknowledge their support.
\end{acknowledgments}

\appendix

\section{Expansion coefficients properties}
\label{apx:cl}

\begin{figure}
	\includegraphics[width=\linewidth]{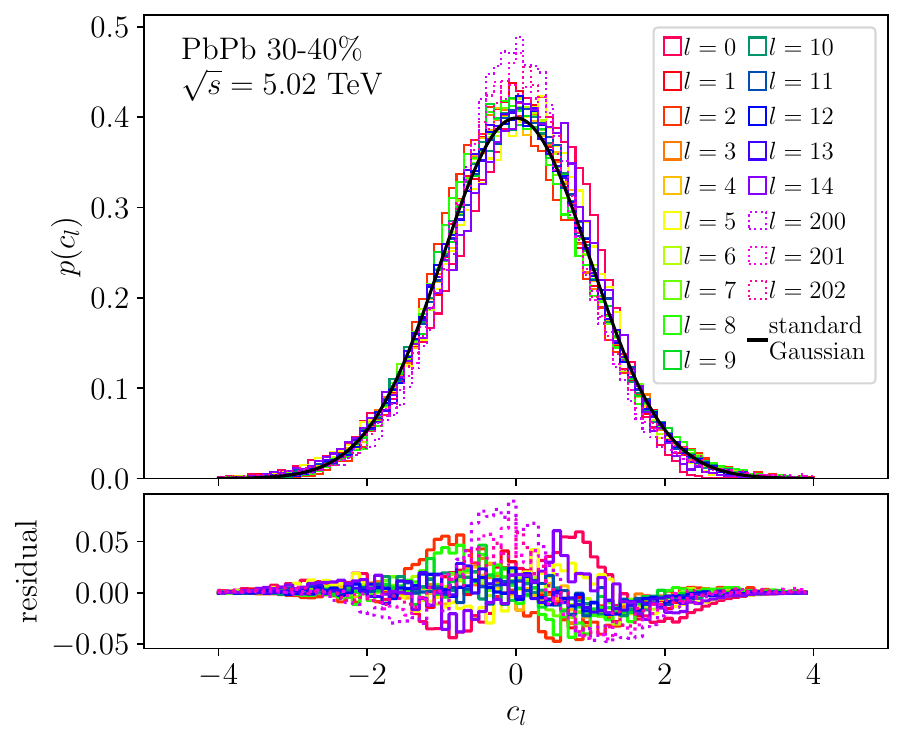}
	\caption{Relative frequency of the expansion coefficients $c_l$ for a few modes (histograms) computed from $2^{15}$ events in the 30--40\% centrality class, compared with a standard Gaussian distribution (full black line).}\vspace{-3mm}
	\label{fig:coef}
\end{figure}

In this appendix we present a few results related to the expansion coefficients $\{c_l\}$ over the basis of fluctuation modes. 
For that purpose, we generated $2^{15}$ random events in the 30--40\% centrality class, which we decomposed according to Eq.~\eqref{eq:decomposition}.

\begin{figure}
	\includegraphics[width=\linewidth]{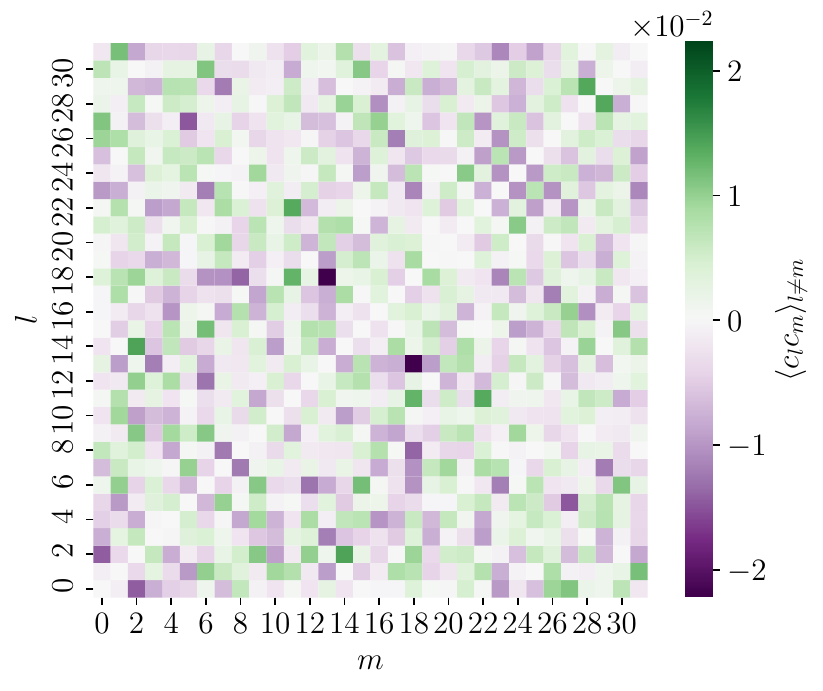}\vspace{-3mm}
	\caption{Cross-correlation $\expval{c_l c_m}$ of the expansion coefficients over two different modes events in the 30--40\% centrality class. The terms on the diagonal are approximately equal to 1 by construction and not shown.}\vspace{-3mm}
	\label{fig:<cl.cm>}
\end{figure}

We first display in Fig.~\ref{fig:coef} the distribution of the $\{c_l\}$ for the leading modes (up to $l=14$) and three higher modes ($l=200$, 201 and 202). 
This shows that each $c_l$ is to a good approximation Gaussian-distributed --- which can be exploited~\cite{Borghini:2022iym} to make further predictions on the distribution of observables --- with a vanishing mean value and a unit variance, where the last two properties should hold by construction, see Eqs.~\eqref{eq:<c_l>=0} and \eqref{eq:<c_l.c_l'>_2} with $l=l'$. 
Some of the distributions present a sizable skewness, also visible on the residual in the bottom panel of the figure.
This especially holds for the leading mode with $l=0$, which is negatively skewed. 
Since this mode $\Psi_{l=0}$ yields negative contributions to the energy per unit space-time rapidity and thus to the multiplicity, see Fig.~\ref{fig:L_IS+FS_30}, the excess of values $c_0<0$ corresponding to the negative skewness signals a tendency to contribute more multiplicity than what the average state alone produces. 
This seems to us to be in agreement with the observation that, due to the fact that larger impact parameters occur more often than smaller ones in the centrality class, the average state is ``more peripheral'' and thus has a smaller multiplicity than the typical initial state in the bin. 
The skewed distribution of $c_0$ then tends to compensate this bias and to produce events with a larger multiplicity. 

In Fig.~\ref{fig:<cl.cm>} we show the cross-correlation $\expval{c_l c_m}_{l\neq m}$ of the expansion coefficients along different fluctuation modes.
Since the average value of each $c_l$ is not exactly zero --- but it is smaller than 0.01 in absolute value ---, these averages do not exactly coincide with the respective covariances. 
Yet the important point is that they all are of order $10^{-2}$, i.e.\ much smaller than the averages $\expval{c_l^2}\simeq 1$, which is a nice check that the fluctuation modes are uncorrelated, see Eq.~\eqref{eq:<c_l.c_l'>_2}.

\section{Further fluctuation modes}
\label{apx:modes2}

\begin{figure*}
	\includegraphics[width=.85\linewidth]{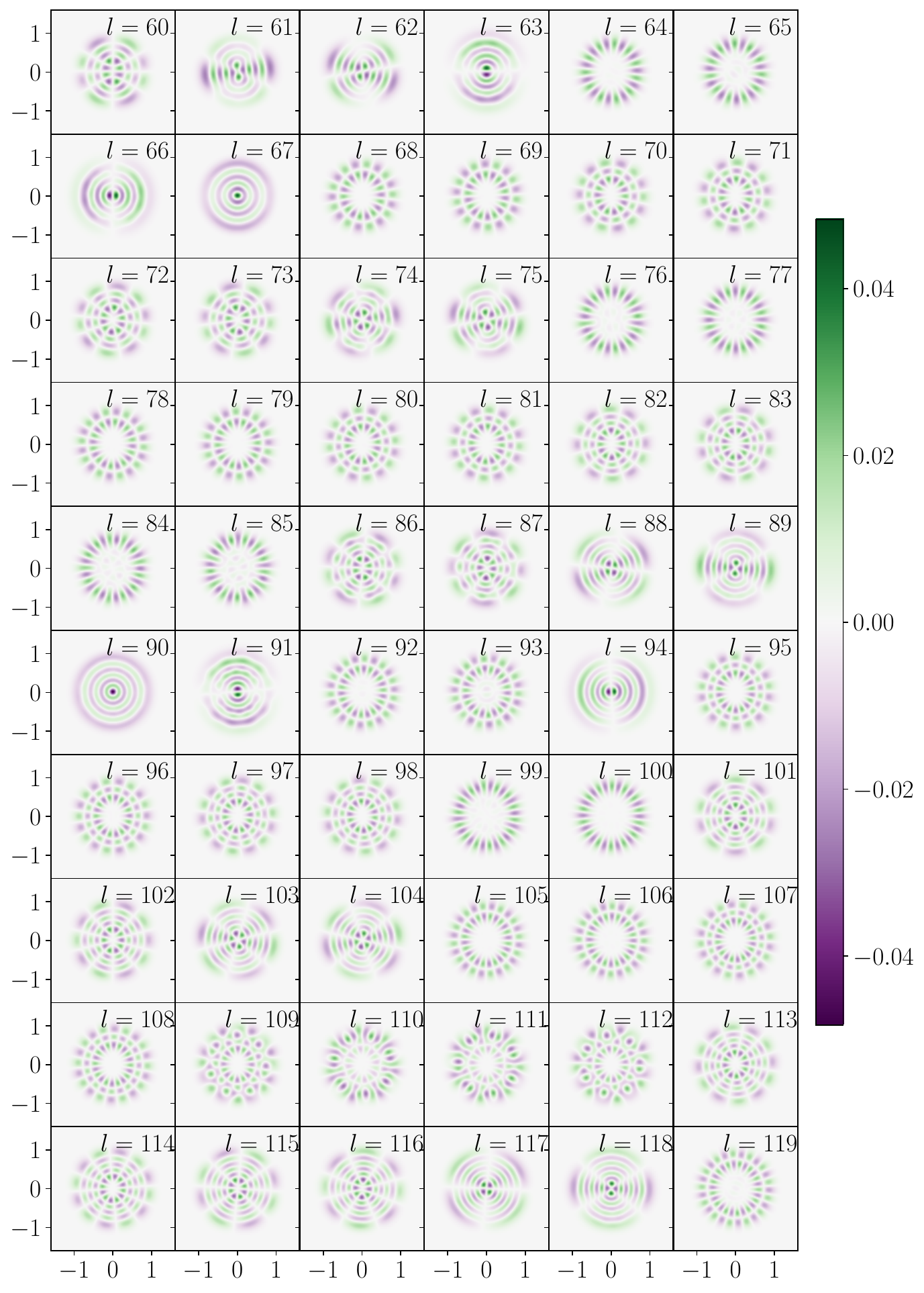}\vspace{-3mm}
	\caption{Normalized transverse profile of the modes $\{\Psi_l\}$ with $60\leq l\leq 119$ for central events. Both axes are in units of the half-density radius $R=6.62$~fm.}
	\label{fig:modes0b}
\end{figure*}

\begin{figure*}
	\includegraphics[width=.85\linewidth]{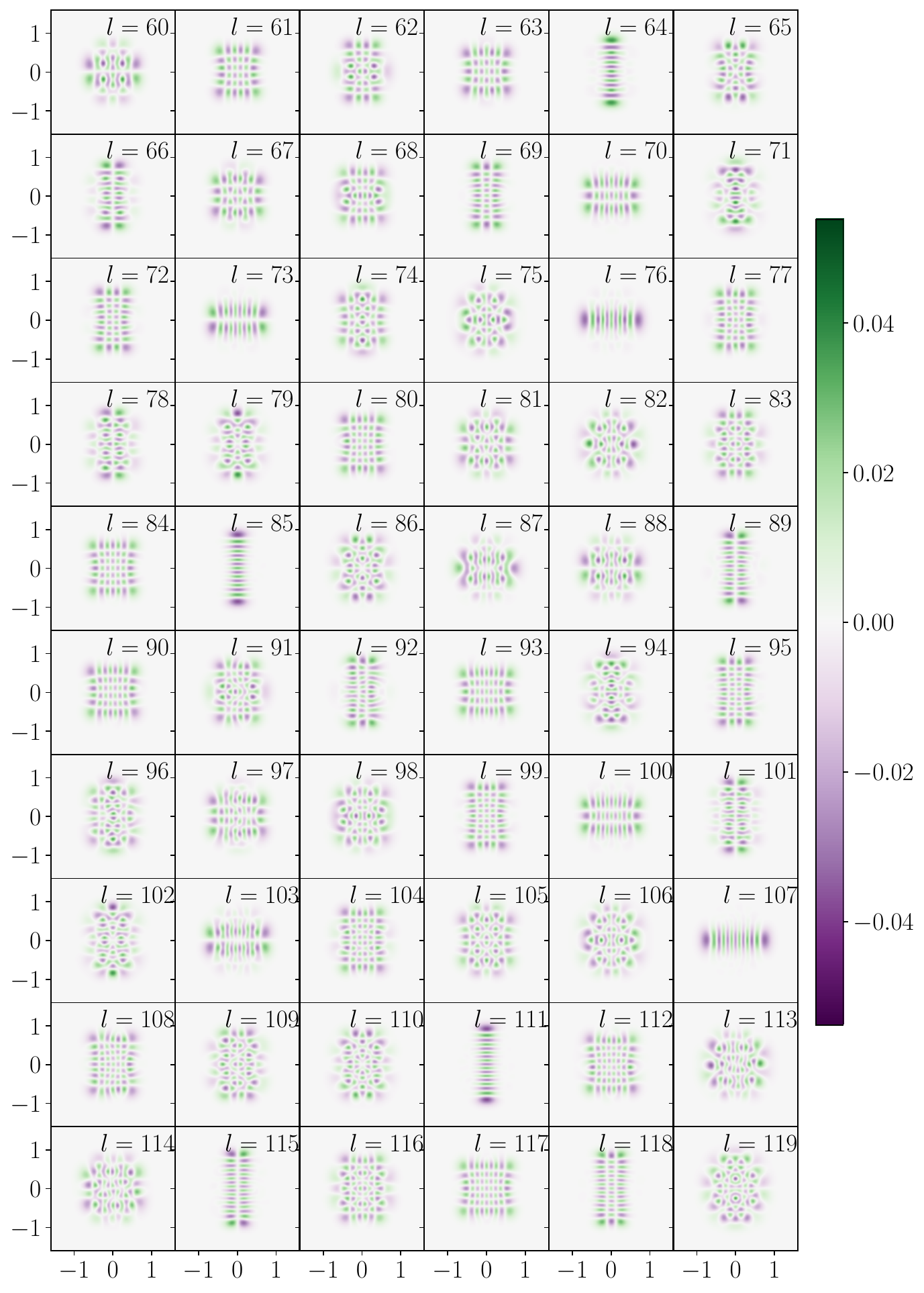}\vspace{-3mm}
	\caption{Normalized transverse profile of the modes $\{\Psi_l\}$ with $60\leq l\leq 119$ for events in the 30--40\% centrality bin. Both axes are in units of the half-density radius $R=6.62$~fm.}
	\label{fig:modes30b}
\end{figure*}

In this appendix we show further normalized eigenvectors of the $\rho$-matrix~\eqref{eq:rho} for the initial state of Pb--Pb collisions at $\sqrt{s_{_{\rm NN}}}=5.02$~TeV within the MC Glauber model. 
Figure~\ref{fig:modes0b} shows the eigenvectors corresponding to the modes $\{\Psi_l\}$ with $60\leq l\leq 119$ in the 0--2.5\% centrality class, complementing the first 60 eigenvectors of Fig.~\ref{fig:modes0}. 
In turn, the eigenvectors corresponding to the modes $\{\Psi_l\}$ with $60\leq l\leq 119$ in the 30--40\% centrality bin are shown in Fig.~\ref{fig:modes30b}, following the eigenvectors of Fig.~\ref{fig:modes30}. 

These eigenvectors show that the trends observed on the first 60 modes, for either centrality class, still persist: 
The fluctuation modes in central collisions (Fig.~\ref{fig:modes0b}) are still roughly circular, with easily recognizable symmetry patterns. 
In contrast, the modes in the 30--40\% centrality class (Fig.~\ref{fig:modes30b}) are more involved.
Note that we know from Fig.~\ref{fig:weights} that these ``higher'' modes in the mid-peripheral events actually have significantly less relative weight than the leading ones.

\section{Optical Glauber model}
\label{apx:optical-Glauber}

\begin{figure}
	\includegraphics[width=\linewidth]{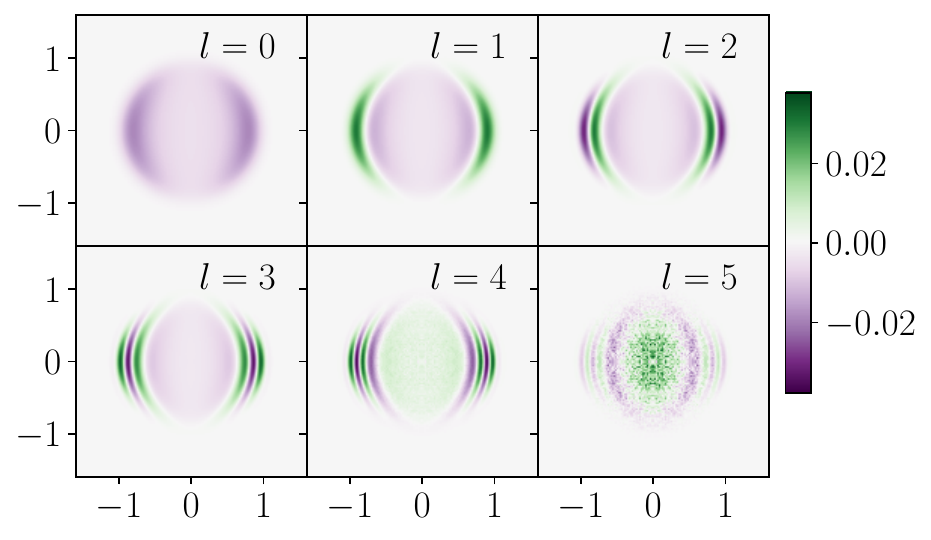}\vspace{-3mm}
	\caption{Normalized transverse profile of the modes $\{\Psi_l^{\rm opt.}\}$ with $0\leq l\leq 5$ for central events ($0\leq b\leq 2.02$~fm) in the optical Glauber model. Both axes are in units of the half-density radius $R=6.62$~fm.}
	\label{fig:optical_modes_0-2.5}
\end{figure}

\begin{figure}
	\includegraphics[width=\linewidth]{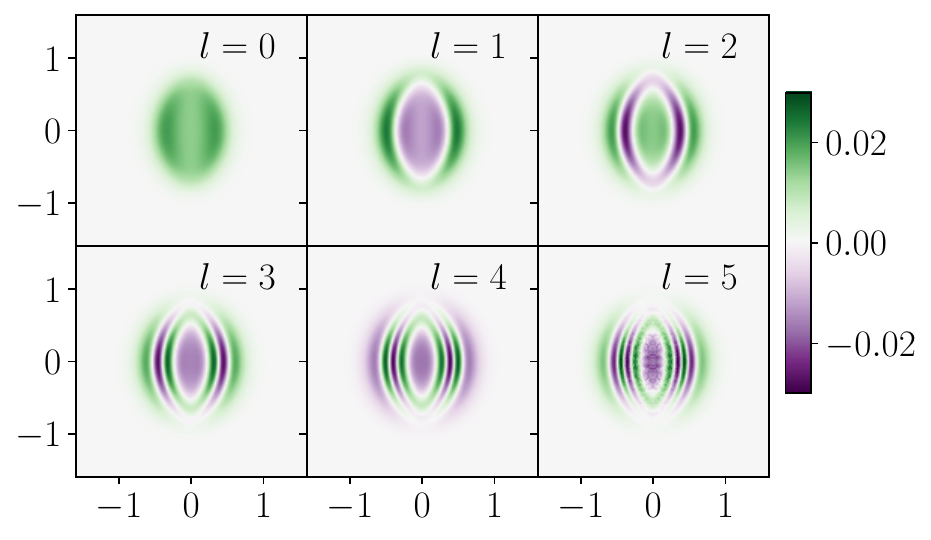}\vspace{-3mm}
	\caption{Same as Fig.~\ref{fig:optical_modes_0-2.5} for events with $7.25\leq b\leq 8.35$~fm.}
	\label{fig:optical_modes_30-40}
\end{figure}

To obtain some idea of the influence of ``geometric'' fluctuations, from the variation of the impact-parameter value ($b$), on the modes of the MC Glauber initial states, we investigated an optical Glauber model with the same underlying Woods--Saxon distribution for the colliding Pb nuclei and the same discretization grid with $\Npts = 192\times 192$ points. 
The amount of energy (density) deposited at a point of the transverse plane was taken proportional to the overlap function of the thickness functions of the two nuclei.\footnote{Since we do not perform any dynamical evolution starting from the initial states of the optical Glauber model, the overall normalization of $e(\bm{x})$ is irrelevant.}
From this energy density $e(\bm{x})$, we determined centrality bins, using the same estimator formula of Ref.~\cite{Giacalone:2019ldn} as for the MC Glauber initial states.
Obviously, the bins are in one-to-one correspondence with ranges in impact parameter: $b\leq 2.02$~fm for the 0--2.5\% centrality class, and $7.25\leq b\leq 8.35$~fm for the 30--40\% bin.

In each centrality class, we generated 1000 initial profiles, from which we computed the average state $\bar{\Psi}^{\rm opt.}$ and the fluctuation modes $\{\Psi^{\rm opt.}_l\}$.
Thanks to our use of the same Woods--Saxon distribution, the average initial state $\bar{\Psi}^{\rm opt.}$ for a given class is very similar to that, shown in Fig.~\ref{fig:avg}, in the MC Glauber model. 
There is actually a small difference: 
on the one hand, using only the overlap of the thickness functions for defining $e(\bm{x})$ means that we do not have the two components of Eq.~\eqref{eq:e(x)}. 
On the other hand, there is no  equivalent in the optical Glauber model of the Gaussian distributions with a width of 0.4~fm used for local energy deposition in the MC Glauber initial states, so that $\bar{\Psi}^{\rm opt.}$ is slightly larger than $\bar{\Psi}$.

More relevant for our purpose in this appendix are the fluctuation modes $\{\Psi^{\rm opt.}_l\}$. 
In either centrality class, only 6 of them --- those with the largest 6 eigenvalues --- have a recognizable shape, the remaining ones are (to the precision of our calculation) only noise and have a relative weight $w_l$ of order $10^{-8}$ or smaller. 
Thus we only show the first 6 normalized eigenvectors corresponding to the $\{\Psi^{\rm opt.}_l\}$ in Figs.~\ref{fig:optical_modes_0-2.5} (for central events) and \ref{fig:optical_modes_30-40} (for events in the 30-40\% centrality class).

At both centralities, the average initial state $\bar{\Psi}^{\rm opt.}$ has a large relative weight $\bar{w}$: 0.89 in central collisions, 0.68 in the 30-40\% centrality bin. 
Then the leading fluctuation mode $\Psi^{\rm opt.}_{l=0}$ carries almost all the rest of the relative weight ($w_0\simeq 0.10$ in the central bin, 0.31 in the midperipheral one). 
Thus the rest carries less than 1\%. 

Thanks to the simple origin of ``fluctuations'' in the optical Glauber model, one can even qualitatively understand roughly the profiles of the first fluctuation modes. 
The key element is that, due to the sampling weight of the impact-parameter value proportional to $b\,\dd b$ and the monotonic relationship between $b$ and multiplicity, the average initial state $\bar{\Psi}^{\rm opt.}$ in a centrality bin $b_{\min}\leq b\leq b_{\max}$ will have a profile more akin to those of the initial states with a value of $b$ closer to $b_{\max}$ than to $b_{\min}$.
Thus, there are events --- the most central ones of the bin, with an impact parameter close to $b_{\min}$ --- which will be broader than $\bar{\Psi}^{\rm opt.}$ with a larger energy density everywhere. 
To account for those ``energetic'' events, it seems almost necessary to have a fluctuation mode with a constant-sign energy density,\footnote{Remember that the sign of the eigenvectors has no intrinsic meaning.} and a significant energy content, as are the leading modes $\Psi^{\rm opt.}_{l=0}$. 
Indeed, one finds that their relative energy content $|\tilde{\varepsilon}_0|_l$ as defined by Eq.~\eqref{eq:mode_eccentricities1} is about 2.6\% in the central bin and 10.8\% in the 30-40\% centrality bin. 

In turn, to reproduce the events with an impact parameter close to the upper value $b_{\max}$ in a centrality bin, which are thus ``more peripheral'' than $\bar{\Psi}^{\rm opt.}$, it is necessary to have less energy density everywhere --- with $e(\bm{x})$ still remaining non-negative!
This gives rise to the modes (with changing sign, such that they are orthogonal to $\Psi^{\rm opt.}_{l=0}$) with $l\geq 1$ of Figs.~\ref{fig:optical_modes_0-2.5} or \ref{fig:optical_modes_30-40}.

Assuming now that the salient features of the optical-Glauber modes $\{\Psi^{\rm opt.}_l\}$ should survive in the modes $\{\Psi_l\}$ of the MC Glauber initial states, the large relative energy content of the leading mode $\Psi^{\rm opt.}_{l=0}$ in the 30--40\% centrality class would ``explain'' the large $|\tilde{\varepsilon}_0|_l$ value of the modes $\Psi_{l=0}$, $\Psi_{l=2}$ and $\Psi_{l=5}$.%
  \footnote{Note that there is no obvious reason why a fluctuation mode $\Psi^{\rm opt.}_l$ of the optical Glauber model should be ``reflected'' in a single mode $\Psi_l$ of the MC Glauber model. 
This is especially true since the range in impact parameter, and the distribution of $b$ within this range, in the midperipheral 30--40\% centrality class of the MC model do not exactly match the range and distribution of the optical Glauber model.}
In turn, the profile of the mode $\Psi^{\rm opt.}_{l=1}$ also in the 30--40\% centrality bin may (partly) explain the shapes of the MC Glauber modes $\Psi_{l=2}$ and $\Psi_{l=5}$ in Fig.~\ref{fig:modes30}, which as we already noted are new compared to the fluctuation modes at fixed $b$ in Ref.~\cite{Borghini:2022iym}, even if this seems to us slightly more speculative than the possible impact of $\Psi^{\rm opt.}_{l=0}$ just mentioned. 

\begin{figure}[h!]
	\includegraphics[width=.485\linewidth]{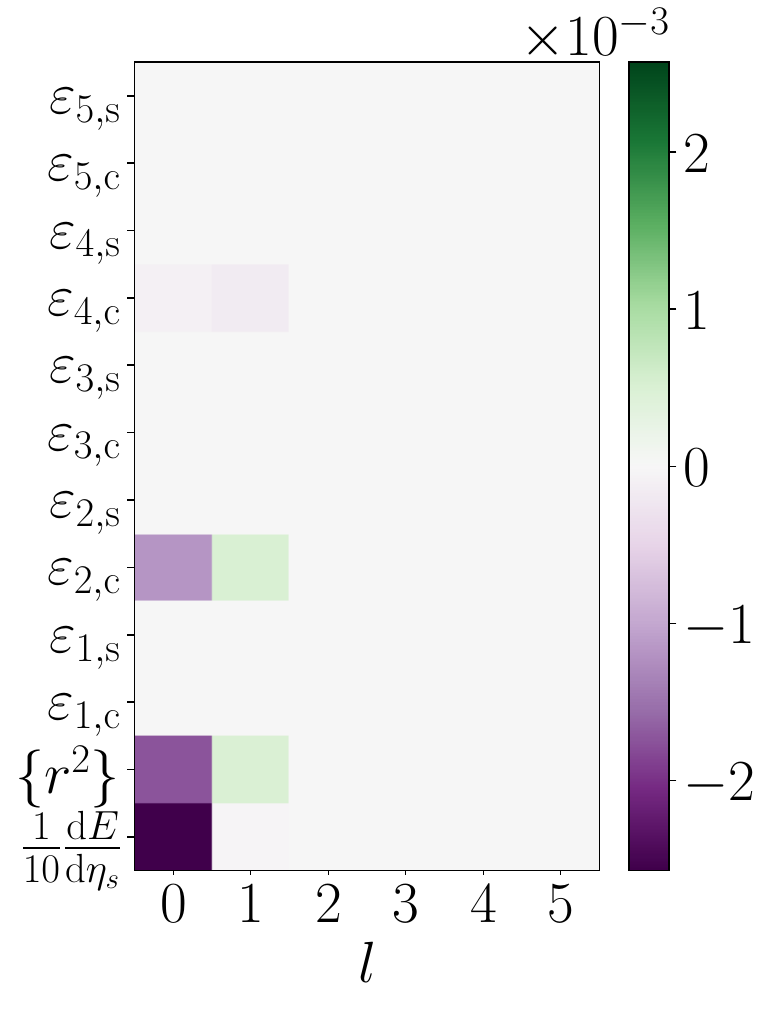}%
	\includegraphics[width=.514\linewidth]{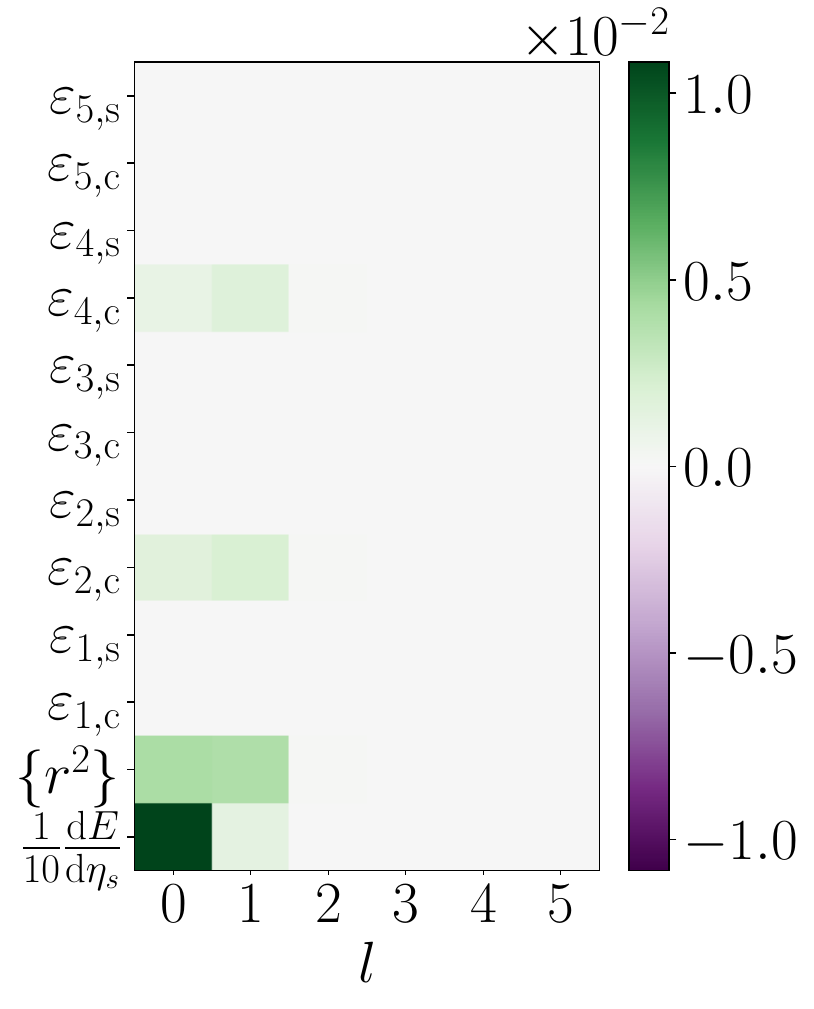}\vspace{-3mm}
	\caption{\label{fig:optical_La}Linear-response coefficients $L_{\alpha,l}$ for initial-state characteristics in the optical Glauber model for events in the 0--2.5\% (left) and 30--40\% (right) centrality classes. The coefficients for $\{r^2\}$ and $\dd E/\dd\eta_s$ have been divided by $\bar{O}_\alpha$.}
\end{figure}

To end up this appendix, we present in Fig.~\ref{fig:optical_La} the linear-response coefficients for the same initial-state observables as in Sec.~\ref{sec:IS}. 
Due to the steeply decreasing relative weights of the fluctuation modes $\{\Psi^{\rm opt.}_l\}$, only the first two ones contribute significantly to observables, namely to the energy density --- as was already discussed above ---, the mean square radius, and the eccentricities $\varepsilon_{2,\rm c}$ and  $\varepsilon_{4,\rm c}$. 
That the other eccentricities vanish could be anticipated, since the optical-Glauber initial states are strictly invariant under the $x\to -x$ and $y\to -y$ (or equivalently $\theta\to\pi-\theta$ and $\theta\to-\theta$) symmetries.

%\begin{figure*}
%    \centering
%    \includegraphics[scale=0.5]{figures/MomentsMode_0-2.5.pdf}
%    \caption{Average $\mu$, variance $\sigma^2$ subtracted by 1, skewness $\gamma_1$ and kurtosis $\gamma_2$ of the probability distribution of the coefficients $c_l$, shown in Fig.~\ref{fig:coef}, for central events $0-2.5\%$.}
%    \label{fig:moments0}
%\end{figure*}
%\begin{figure*}
%    \centering
%    \includegraphics[scale=0.5]{figures/MomentsMode_30-40.pdf}
%    \caption{The same plot as in Fig~\ref{fig:moments0}, but for non-central events $30-40\%$.}
%    \label{fig:moments30}
%\end{figure*}

%\bibliography{ref.bib}%

\begin{thebibliography}{99}

\bibitem{Chaudhuri:2012yt}
A.~K.~Chaudhuri,
{\it A short course on Relativistic Heavy Ion Collisions\/},
(IOPP, Bristol, 2014)
%ISBN 978-0-7503-1061-1, 978-0-7503-1060-4
%doi:10.1088/978-0-750-31060-4
arXiv:1207.7028 [nucl-th]

\bibitem{Elfner:2022iae}
H.~Elfner and B.~M{\"u}ller,
%``The exploration of hot and dense nuclear matter: introduction to relativistic heavy-ion physics,''
J. Phys. G \textbf{50}, 103001 (2023).
arXiv:2210.12056 [nucl-th]

\bibitem{Drescher:2006ca}
H.~J.~Drescher and Y.~Nara,
%``Effects of fluctuations on the initial eccentricity from the Color Glass Condensate in heavy ion collisions,''
Phys. Rev. C \textbf{75}, 034905 (2007).
%doi:10.1103/PhysRevC.75.034905
arXiv:nucl-th/0611017 [nucl-th]

%\cite{Miller:2007ri}
\bibitem{Miller:2007ri}
M.~L.~Miller, K.~Reygers, S.~J.~Sanders and P.~Steinberg,
%``Glauber modeling in high energy nuclear collisions,''
Ann. Rev. Nucl. Part. Sci. \textbf{57}, 205 (2007).
%doi:10.1146/annurev.nucl.57.090506.123020
arXiv:nucl-ex/0701025 [nucl-ex]

\bibitem{Broniowski:2007nz}
W.~Broniowski, M.~Rybczynski and P.~Bo\.zek,
%``GLISSANDO: Glauber initial-state simulation and more..,''
Comput. Phys. Commun. \textbf{180}, 69 (2009).
%doi:10.1016/j.cpc.2008.07.016
arXiv:0710.5731 [nucl-th]

\bibitem{Albacete:2014fwa}
J.~L.~Albacete and C.~Marquet,
%``Gluon saturation and initial conditions for relativistic heavy ion collisions,''
Prog. Part. Nucl. Phys. \textbf{76}, 1 (2014).
%doi:10.1016/j.ppnp.2014.01.004
arXiv:1401.4866 [hep-ph]

\bibitem{Loizides:2014vua}
C.~Loizides, J.~Nagle and P.~Steinberg,
%``Improved version of the PHOBOS Glauber Monte Carlo,''
SoftwareX \textbf{1-2}, 13 (2015).
%doi:10.1016/j.softx.2015.05.001
arXiv:1408.2549 [nucl-ex]

\bibitem{Moreland:2014oya}
J.~S.~Moreland, J.~E.~Bernhard and S.~A.~Bass,
%``Alternative ansatz to wounded nucleon and binary collision scaling in high-energy nuclear collisions,''
Phys. Rev. C \textbf{92}, 011901 (2015).
%doi:10.1103/PhysRevC.92.011901
arXiv:1412.4708 [nucl-th]

\bibitem{Niemi:2015qia}
H.~Niemi, K.~J.~Eskola and R.~Paatelainen,
%``Event-by-event fluctuations in a perturbative QCD + saturation + hydrodynamics model: Determining QCD matter shear viscosity in ultrarelativistic heavy-ion collisions,''
Phys. Rev. C \textbf{93}, 024907 (2016).
%doi:10.1103/PhysRevC.93.024907
arXiv:1505.02677 [hep-ph]

\bibitem{Shen:2017bsr}
C.~Shen and B.~Schenke,
%``Dynamical initial state model for relativistic heavy-ion collisions,''
Phys. Rev. C \textbf{97}, 024907 (2018).
%doi:10.1103/PhysRevC.97.024907
arXiv:1710.00881 [nucl-th]

\bibitem{Giacalone:2019kgg}
G.~Giacalone, P.~Guerrero-Rodr{\'\i}guez, M.~Luzum, C.~Marquet and J.-Y.~Ollitrault,
%``Fluctuations in heavy-ion collisions generated by QCD interactions in the color glass condensate effective theory,''
Phys. Rev. C \textbf{100}, 024905 (2019).
%doi:10.1103/PhysRevC.100.024905
arXiv:1902.07168 [nucl-th]

\bibitem{Soeder:2023vdn}
D.~Soeder, W.~Ke, J.~F.~Paquet and S.~A.~Bass,
%``Bayesian parameter estimation with a new three-dimensional initial-conditions model for ultrarelativistic heavy-ion collisions,''
arXiv:2306.08665 [nucl-th]

\bibitem{Garcia-Montero:2023gex}
O.~Garcia-Montero, H.~Elfner and S.~Schlichting,
%``McDIPPER: A novel saturation-based 3+1D initial-state model for heavy ion collisions,''
Phys. Rev. C \textbf{109}, 044916 (2024).
%doi:10.1103/PhysRevC.109.044916
arXiv:2308.11713 [hep-ph]

\bibitem{Kuha:2024kmq}
M.~Kuha, J.~Auvinen, K.~J.~Eskola, H.~Hirvonen, Y.~Kanakubo and H.~Niemi,
%``Monte Carlo event generator MC-EKRT with saturated minijet production for initializing (3+1)D fluid dynamics in high-energy nuclear collisions,''
Phys. Rev. C \textbf{111}, 054914 (2025).
%doi:10.1103/PhysRevC.111.054914
arXiv:2406.17592 [hep-ph]

\bibitem{Socolowski:2004hw}
O.~Socolowski, Jr., F.~Grassi, Y.~Hama and T.~Kodama,
%``Fluctuations of the initial conditions and the continuous emission in hydro description of two-pion interferometry,''
Phys. Rev. Lett. \textbf{93}, 182301 (2004).
%doi:10.1103/PhysRevLett.93.182301
arXiv:hep-ph/0405181 [hep-ph]

\bibitem{Broniowski:2007ft}
W.~Broniowski, P.~Bo\.zek and M.~Rybczynski,
%``Fluctuating initial conditions in heavy-ion collisions from the Glauber approach,''
Phys. Rev. C \textbf{76}, 054905 (2007).
%doi:10.1103/PhysRevC.76.054905
arXiv:0706.4266 [nucl-th]

\bibitem{Schenke:2012wb}
B.~Schenke, P.~Tribedy and R.~Venugopalan,
%``Fluctuating Glasma initial conditions and flow in heavy ion collisions,''
Phys. Rev. Lett. \textbf{108}, 252301 (2012).
%doi:10.1103/PhysRevLett.108.252301
arXiv:1202.6646 [nucl-th]

\bibitem{Schenke:2012hg}
B.~Schenke, P.~Tribedy and R.~Venugopalan,
%``Event-by-event gluon multiplicity, energy density, and eccentricities in ultrarelativistic heavy-ion collisions,''
Phys. Rev. C \textbf{86}, 034908 (2012).
%doi:10.1103/PhysRevC.86.034908
arXiv:1206.6805 [hep-ph]

\bibitem{Paatelainen:2012at}
R.~Paatelainen, K.~J.~Eskola, H.~Holopainen and K.~Tuominen,
%``Multiplicities and $p_T$ spectra in ultrarelativistic heavy ion collisions from a next-to-leading order improved perturbative QCD + saturation + hydrodynamics model,''
Phys. Rev. C \textbf{87}, 044904 (2013).
%doi:10.1103/PhysRevC.87.044904
arXiv:1211.0461 [hep-ph]

\bibitem{Liu:2015nwa}
J.~Liu, C.~Shen and U.~Heinz,
%``Pre-equilibrium evolution effects on heavy-ion collision observables,''
Phys. Rev. C \textbf{91}, 064906 (2015).
%doi:10.1103/PhysRevC.91.064906
arXiv:1504.02160 [nucl-th]
[Erratum: Phys. Rev. C \textbf{92}, 049904 (2015)]

\bibitem{Shen:2020jwv}
C.~Shen and S.~Alzhrani,
%``Collision-geometry-based 3D initial condition for relativistic heavy-ion collisions,''
Phys. Rev. C \textbf{102}, 014909 (2020).
%doi:10.1103/PhysRevC.102.014909
arXiv:2003.05852 [nucl-th]

\bibitem{NunesdaSilva:2020bfs}
T.~Nunes da Silva, D.~Chinellato, M.~Hippert, W.~Serenone, J.~Takahashi, G.~S.~Denicol, M.~Luzum and J.~Noronha,
%``Pre-hydrodynamic evolution and its signatures in final-state heavy-ion observables,''
Phys. Rev. C \textbf{103}, 054906 (2021).
%doi:10.1103/PhysRevC.103.054906
arXiv:2006.02324 [nucl-th]

\bibitem{Shen:2022oyg}
C.~Shen and B.~Schenke,
%``Longitudinal dynamics and particle production in relativistic nuclear collisions,''
Phys. Rev. C \textbf{105}, 064905 (2022).
%doi:10.1103/PhysRevC.105.064905
arXiv:2203.04685 [nucl-th]

\bibitem{Novak:2013bqa}
J.~Novak, K.~Novak, S.~Pratt, J.~Vredevoogd, C.~Coleman-Smith and R.~Wolpert,
%``Determining Fundamental Properties of Matter Created in Ultrarelativistic Heavy-Ion Collisions,''
Phys. Rev. C \textbf{89}, 034917 (2014).
%doi:10.1103/PhysRevC.89.034917
arXiv:1303.5769 [nucl-th]

\bibitem{Pratt:2015zsa}
S.~Pratt, E.~Sangaline, P.~Sorensen and H.~Wang,
%``Constraining the Eq. of State of Super-Hadronic Matter from Heavy-Ion Collisions,''
Phys. Rev. Lett. \textbf{114}, 202301 (2015).
%doi:10.1103/PhysRevLett.114.202301
arXiv:1501.04042 [nucl-th]

\bibitem{Bernhard:2015hxa}
J.~E.~Bernhard, P.~W.~Marcy, C.~E.~Coleman-Smith, S.~Huzurbazar, R.~L.~Wolpert and S.~A.~Bass,
%``Quantifying properties of hot and dense QCD matter through systematic model-to-data comparison,''
Phys. Rev. C \textbf{91}, 054910 (2015).
%doi:10.1103/PhysRevC.91.054910
arXiv:1502.00339 [nucl-th]

\bibitem{Bernhard:2016tnd}
J.~E.~Bernhard, J.~S.~Moreland, S.~A.~Bass, J.~Liu and U.~Heinz,
%``Applying Bayesian parameter estimation to relativistic heavy-ion collisions: simultaneous characterization of the initial state and quark-gluon plasma medium,''
Phys. Rev. C \textbf{94}, 024907 (2016).
%doi:10.1103/PhysRevC.94.024907
arXiv:1605.03954 [nucl-th]

\bibitem{Auvinen:2017fjw}
J.~Auvinen, J.~E.~Bernhard, S.~A.~Bass and I.~Karpenko,
%``Investigating the collision energy dependence of $\eta$/s in the beam energy scan at the BNL Relativistic Heavy Ion Collider using Bayesian statistics,''
Phys. Rev. C \textbf{97}, 044905 (2018).
%doi:10.1103/PhysRevC.97.044905
arXiv:1706.03666 [hep-ph]

\bibitem{Bernhard:2019bmu}
J.~E.~Bernhard, J.~S.~Moreland and S.~A.~Bass,
%``Bayesian estimation of the specific shear and bulk viscosity of quark{\textendash}gluon plasma,''
Nature Phys. \textbf{15}, 1113 (2019).
%doi:10.1038/s41567-019-0611-8

\bibitem{Moreland:2018gsh}
J.~S.~Moreland, J.~E.~Bernhard and S.~A.~Bass,
%``Bayesian calibration of a hybrid nuclear collision model using p-Pb and Pb-Pb data at energies available at the CERN Large Hadron Collider,''
Phys. Rev. C \textbf{101}, 024911 (2020).
%doi:10.1103/PhysRevC.101.024911
arXiv:1808.02106 [nucl-th]

\bibitem{JETSCAPE:2020avt}
J.~F.~Paquet \textit{et al.} [JETSCAPE Collaboration],
%``Revisiting Bayesian constraints on the transport coefficients of QCD,''
Nucl. Phys. A \textbf{1005}, 121749 (2021).
%doi:10.1016/j.nuclphysa.2020.121749
arXiv:2002.05337 [nucl-th]

\bibitem{JETSCAPE:2020shq}
D.~Everett \textit{et al.} [JETSCAPE Collaboration],
%``Phenomenological constraints on the transport properties of QCD matter with data-driven model averaging,''
Phys. Rev. Lett. \textbf{126}, 242301 (2021).
%doi:10.1103/PhysRevLett.126.242301
arXiv:2010.03928 [hep-ph]

\bibitem{Nijs:2020ors}
G.~Nijs, W.~van der Schee, U.~G{\"u}rsoy and R.~Snellings,
%``Transverse Momentum Differential Global Analysis of Heavy-Ion Collisions,''
Phys. Rev. Lett. \textbf{126}, 202301 (2021).
%doi:10.1103/PhysRevLett.126.202301
arXiv:2010.15130 [nucl-th]

\bibitem{Nijs:2020roc}
G.~Nijs, W.~van der Schee, U.~G{\"u}rsoy and R.~Snellings,
%``Bayesian analysis of heavy ion collisions with the heavy ion computational framework Trajectum,''
Phys. Rev. C \textbf{103}, 054909 (2021).
%doi:10.1103/PhysRevC.103.054909
arXiv:2010.15134 [nucl-th]

\bibitem{JETSCAPE:2020mzn}
D.~Everett \textit{et al.} [JETSCAPE Collaboration],
%``Multisystem Bayesian constraints on the transport coefficients of QCD matter,''
Phys. Rev. C \textbf{103}, 054904 (2021).
%doi:10.1103/PhysRevC.103.054904
arXiv:2011.01430 [hep-ph]

\bibitem{JETSCAPE:2021ehl}
S.~Cao \textit{et al.} [JETSCAPE Collaboration],
%``Determining the jet transport coefficient q̂ from inclusive hadron suppression measurements using Bayesian parameter estimation,''
Phys. Rev. C \textbf{104}, 024905 (2021).
%doi:10.1103/PhysRevC.104.024905
arXiv:2102.11337 [nucl-th]

\bibitem{Parkkila:2021tqq}
J.~E.~Parkkila, A.~Onnerstad and D.~J.~Kim,
%``Bayesian estimation of the specific shear and bulk viscosity of the quark-gluon plasma with additional flow harmonic observables,''
Phys. Rev. C \textbf{104}, 054904 (2021).
%doi:10.1103/PhysRevC.104.054904
arXiv:2106.05019 [hep-ph]

\bibitem{Parkkila:2021yha}
J.~E.~Parkkila, A.~Onnerstad, S.~F.~Taghavi, C.~Mordasini, A.~Bilandzic, M.~Virta and D.~J.~Kim,
%``New constraints for QCD matter from improved Bayesian parameter estimation in heavy-ion collisions at LHC,''
Phys. Lett. B \textbf{835}, 137485 (2022).
%doi:10.1016/j.physletb.2022.137485
arXiv:2111.08145 [hep-ph]

\bibitem{Xie:2022ght}
M.~Xie, W.~Ke, H.~Zhang and X.~N.~Wang,
%``Information-field-based global Bayesian inference of the jet transport coefficient,''
Phys. Rev. C \textbf{108}, L011901 (2023).
%doi:10.1103/PhysRevC.108.L011901
arXiv:2206.01340 [hep-ph]

\bibitem{Heffernan:2023utr}
M.~R.~Heffernan, C.~Gale, S.~Jeon and J.~F.~Paquet,
%``Bayesian quantification of strongly interacting matter with color glass condensate initial conditions,''
Phys. Rev. C \textbf{109}, 065207 (2024).
%doi:10.1103/PhysRevC.109.065207
arXiv:2302.09478 [nucl-th]

\bibitem{Liyanage:2023nds}
D.~Liyanage, {\"O}.~S{\"u}rer, M.~Plumlee, S.~M.~Wild and U.~Heinz,
%``Bayesian calibration of viscous anisotropic hydrodynamic simulations of heavy-ion collisions,''
Phys. Rev. C \textbf{108}, 054905 (2023).
%doi:10.1103/PhysRevC.108.054905
arXiv:2302.14184 [nucl-th]

\bibitem{Heffernan:2023gye}
M.~R.~Heffernan, C.~Gale, S.~Jeon and J.~F.~Paquet,
%``Early-Times Yang-Mills Dynamics and the Characterization of Strongly Interacting Matter with Statistical Learning,''
Phys. Rev. Lett. \textbf{132}, 252301 (2024).
%doi:10.1103/PhysRevLett.132.252301
arXiv:2306.09619 [nucl-th]

\bibitem{Shen:2023awv}
C.~Shen, B.~Schenke and W.~Zhao,
%``Viscosities of the Baryon-Rich Quark-Gluon Plasma from Beam Energy Scan Data,''
Phys. Rev. Lett. \textbf{132}, 072301 (2024).
%doi:10.1103/PhysRevLett.132.072301
arXiv:2310.10787 [nucl-th]

\bibitem{Jahan:2024wpj}
S.~A.~Jahan, H.~Roch and C.~Shen,
%``Bayesian analysis of (3+1)D relativistic nuclear dynamics with the RHIC beam energy scan data,''
Phys. Rev. C \textbf{110}, 054905 (2024).
%doi:10.1103/PhysRevC.110.054905
arXiv:2408.00537 [nucl-th]

\bibitem{JETSCAPE:2024cqe}
R.~Ehlers \textit{et al.} [JETSCAPE Collaboration],
%``Bayesian inference analysis of jet quenching using inclusive jet and hadron suppression measurements,''
Phys. Rev. C \textbf{111}, 054913 (2025).
%doi:10.1103/PhysRevC.111.054913
arXiv:2408.08247 [hep-ph]

\bibitem{Domingues:2024pom}
T.~S.~Domingues, R.~Krupczak, J.~Noronha, T.~N.~da Silva, J.~F.~Paquet and M.~Luzum,
%``Effect of causality constraints on Bayesian analyses of heavy-ion collisions,''
Phys. Rev. C \textbf{110}, 064904 (2024).
%doi:10.1103/PhysRevC.110.064904
arXiv:2409.17127 [nucl-th]

\bibitem{Gotz:2025wnv}
N.~G{\"o}tz, I.~Karpenko and H.~Elfner,
%``Bayesian analysis of a (3+1)D hybrid approach with initial conditions from hadronic transport,''
Phys. Rev. C \textbf{112}, 014910 (2025).
%doi:10.1103/rzml-rjxz
arXiv:2503.10181 [nucl-th]

\bibitem{Jahan:2025cbp}
S.~A.~Jahan, H.~Roch and C.~Shen,
%``Bayesian Model Selection and Uncertainty Propagation for Beam Energy Scan Heavy-Ion Collisions,''
arXiv:2507.11394 [nucl-th]

\bibitem{Bhalerao:2014mua}
R.~S.~Bhalerao, J.-Y.~Ollitrault, S.~Pal and D.~Teaney,
%``Principal component analysis of event-by-event fluctuations,''
Phys. Rev. Lett. \textbf{114}, 152301 (2015).
%doi:10.1103/PhysRevLett.114.152301
arXiv:1410.7739 [nucl-th]

\bibitem{Mazeliauskas:2015vea}
A.~Mazeliauskas and D.~Teaney,
%``Subleading harmonic flows in hydrodynamic simulations of heavy ion collisions,''
Phys. Rev. C \textbf{91}, 044902 (2015).
%doi:10.1103/PhysRevC.91.044902
arXiv:1501.03138 [nucl-th]

\bibitem{Mazeliauskas:2015efa}
A.~Mazeliauskas and D.~Teaney,
%``Fluctuations of harmonic and radial flow in heavy ion collisions with principal components,''
Phys. Rev. C \textbf{93}, 024913 (2016).
%doi:10.1103/PhysRevC.93.024913
arXiv:1509.07492 [nucl-th]

\bibitem{Floerchinger:2013rya}
S.~Floerchinger and U.~A.~Wiedemann,
%``Mode-by-mode fluid dynamics for relativistic heavy ion collisions,''
Phys. Lett. B \textbf{728}, 407 (2014).
%doi:10.1016/j.physletb.2013.12.025
arXiv:1307.3453 [hep-ph]

\bibitem{Floerchinger:2013vua}
S.~Floerchinger and U.~A.~Wiedemann,
%``Characterization of initial fluctuations for the hydrodynamical description of heavy ion collisions,''
Phys. Rev. C \textbf{88}, 044906 (2013).
%doi:10.1103/PhysRevC.88.044906
arXiv:1307.7611 [hep-ph]

\bibitem{Floerchinger:2013hza}
S.~Floerchinger and U.~A.~Wiedemann,
%``Kinetic freeze-out, particle spectra and harmonic flow coefficients from mode-by-mode hydrodynamics,''
Phys. Rev. C \textbf{89}, 034914 (2014).
%doi:10.1103/PhysRevC.89.034914
arXiv:1311.7613 [hep-ph]

\bibitem{Floerchinger:2013tya}
S.~Floerchinger, U.~A.~Wiedemann, A.~Beraudo, L.~Del Zanna, G.~Inghirami and V.~Rolando,
%``How (non-)linear is the hydrodynamics of heavy ion collisions?,''
Phys. Lett. B \textbf{735}, 305 (2014).
%doi:10.1016/j.physletb.2014.06.049
arXiv:1312.5482 [hep-ph]

\bibitem{Floerchinger:2014fta}
S.~Floerchinger and U.~A.~Wiedemann,
%``Statistics of initial density perturbations in heavy ion collisions and their fluid dynamic response,''
JHEP \textbf{08}, 005 (2014).
%doi:10.1007/JHEP08(2014)005
arXiv:1405.4393 [hep-ph]

\bibitem{Floerchinger:2018pje}
S.~Floerchinger, E.~Grossi and J.~Lion,
%``Fluid dynamics of heavy ion collisions with mode expansion,''
Phys. Rev. C \textbf{100}, 014905 (2019).
%doi:10.1103/PhysRevC.100.014905
arXiv:1811.01870 [nucl-th]

\bibitem{Coleman-Smith:2012kbb}
C.~E.~Coleman-Smith, H.~Petersen and R.~L.~Wolpert,
%``Classification of initial state granularity via 2d Fourier Expansion,''
J. Phys. G \textbf{40}, 095103 (2013).
%doi:10.1088/0954-3899/40/9/095103
arXiv:1204.5774 [hep-ph]

\bibitem{Borghini:2022iym}
N.~Borghini, M.~Borrell, N.~Feld, H.~Roch, S.~Schlichting and C.~Werthmann,
%``Statistical analysis of initial-state and final-state response in heavy-ion collisions,''
Phys. Rev. C \textbf{107}, 034905 (2023).
%doi:10.1103/PhysRevC.107.034905
arXiv:2209.01176 [hep-ph]

\bibitem{dEnterria:2020dwq}
D.~d'Enterria and C.~Loizides,
%``Progress in the Glauber Model at Collider Energies,''
Ann. Rev. Nucl. Part. Sci. \textbf{71}, 315 (2021).
%doi:10.1146/annurev-nucl-102419-060007
arXiv:2011.14909 [hep-ph]

\bibitem{Giacalone:2019ldn}
G.~Giacalone, A.~Mazeliauskas and S.~Schlichting,
%``Hydrodynamic attractors, initial state energy and particle production in relativistic nuclear collisions,''
Phys. Rev. Lett. \textbf{123}, 262301 (2019).
%doi:10.1103/PhysRevLett.123.262301
arXiv:1908.02866 [hep-ph]

\bibitem{ALICE:2015juo}
J.~Adam \textit{et al.} [ALICE Collaboration],
%``Centrality Dependence of the Charged-Particle Multiplicity Density at Midrapidity in Pb-Pb Collisions at $\sqrt{s_{\rm NN}}$ = 5.02 TeV,''
Phys. Rev. Lett. \textbf{116}, 222302 (2016).
%doi:10.1103/PhysRevLett.116.222302
arXiv:1512.06104 [nucl-ex]

\bibitem{Roch-Krupczak2024}
H.~Roch, R.~Krupczak, Hendrik1704/EBE-PREEQ-HYDRO-TRANSPORT: v2.1 (2024). Zenodo. https://doi.org/10.5281/zenodo.12694840

\bibitem{Kurkela:2018vqr}
A.~Kurkela, A.~Mazeliauskas, J.~F.~Paquet, S.~Schlichting and D.~Teaney,
%``Effective kinetic description of event-by-event pre-equilibrium dynamics in high-energy heavy-ion collisions,''
Phys. Rev. C \textbf{99}, 034910 (2019).
%doi:10.1103/PhysRevC.99.034910
arXiv:1805.00961 [hep-ph]

\bibitem{Schenke:2010nt}
B.~Schenke, S.~Jeon and C.~Gale,
%``(3+1)D hydrodynamic simulation of relativistic heavy-ion collisions,''
Phys. Rev. C \textbf{82}, 014903 (2010).
%doi:10.1103/PhysRevC.82.014903
arXiv:1004.1408 [hep-ph]

\bibitem{Schenke:2010rr}
B.~Schenke, S.~Jeon and C.~Gale,
%``Elliptic and triangular flow in event-by-event (3+1)D viscous hydrodynamics,''
Phys. Rev. Lett. \textbf{106}, 042301 (2011).
%doi:10.1103/PhysRevLett.106.042301
arXiv:1009.3244 [hep-ph]

\bibitem{Paquet:2015lta}
J.~F.~Paquet, C.~Shen, G.~S.~Denicol, M.~Luzum, B.~Schenke, S.~Jeon and C.~Gale,
%``Production of photons in relativistic heavy-ion collisions,''
Phys. Rev. C \textbf{93}, 044906 (2016).
%doi:10.1103/PhysRevC.93.044906
arXiv:1509.06738 [hep-ph]

\bibitem{Borghini:2024kll}
N.~Borghini, R.~Krupczak and H.~Roch,
%``Comparing matching prescriptions between pre-equilibrium and hydrodynamic models in high-energy nuclear collisions,''
Eur. Phys. J. C \textbf{84}, 1128 (2024).
%doi:10.1140/epjc/s10052-024-13509-8
arXiv:2407.10634 [nucl-th]

\bibitem{HotQCD:2014kol}
A.~Bazavov \textit{et al.} [HotQCD Collaboration],
%``Equation of state in ( 2+1 )-flavor QCD,''
Phys. Rev. D \textbf{90}, 094503 (2014).
%doi:10.1103/PhysRevD.90.094503
arXiv:1407.6387 [hep-lat]

\bibitem{Bernhard:2018hnz}
J.~E.~Bernhard,
%``Bayesian parameter estimation for relativistic heavy-ion collisions,''
arXiv:1804.06469 [nucl-th]

\bibitem{Shen:2014vra}
C.~Shen, Z.~Qiu, H.~Song, J.~Bernhard, S.~Bass and U.~Heinz,
%``The iEBE-VISHNU code package for relativistic heavy-ion collisions,''
Comput. Phys. Commun. \textbf{199}, 61 (2016).
%doi:10.1016/j.cpc.2015.08.039
arXiv:1409.8164 [nucl-th]

\bibitem{SMASH:2016zqf}
J.~Weil \textit{et al.} [SMASH Collaboration],
%``Particle production and equilibrium properties within a new hadron transport approach for heavy-ion collisions,''
Phys. Rev. C \textbf{94}, 054905 (2016).
%doi:10.1103/PhysRevC.94.054905
arXiv:1606.06642 [nucl-th]

\bibitem{Borghini:2024ekn}
N.~Borghini, H.~Roch and A.~Sch{\"u}tte,
%``Statistical analysis of the fluctuations of an initial-state model with independently distributed hot spots,''
Eur. Phys. J. C \textbf{85}, 12 (2025).
%doi:10.1140/epjc/s10052-024-13694-6
arXiv:2402.07888 [nucl-th]

\bibitem{Teaney:2010vd}
D.~Teaney and L.~Yan,
%``Triangularity and Dipole Asymmetry in Heavy Ion Collisions,''
Phys. Rev. C \textbf{83}, 064904 (2011).
%doi:10.1103/PhysRevC.83.064904
arXiv:1010.1876 [nucl-th]

\bibitem{Gardim:2011xv}
F.~G.~Gardim, F.~Grassi, M.~Luzum and J.-Y.~Ollitrault,
%``Mapping the hydrodynamic response to the initial geometry in heavy-ion collisions,''
Phys. Rev. C \textbf{85}, 024908 (2012).
%doi:10.1103/PhysRevC.85.024908
arXiv:1111.6538 [nucl-th]

\bibitem{Borghini:2005kd}
N.~Borghini and J.-Y.~Ollitrault,
%``Momentum spectra, anisotropic flow, and ideal fluids,''
Phys. Lett. B \textbf{642}, 227 (2006).
%doi:10.1016/j.physletb.2006.09.062
arXiv:nucl-th/0506045 [nucl-th]

\bibitem{Teaney:2012ke}
D.~Teaney and L.~Yan,
%``Non linearities in the harmonic spectrum of heavy ion collisions with ideal and viscous hydrodynamics,''
Phys. Rev. C \textbf{86} (2012), 044908.
%doi:10.1103/PhysRevC.86.044908
arXiv:1206.1905 [nucl-th]

\bibitem{Niemi:2012aj}
H.~Niemi, G.~S.~Denicol, H.~Holopainen and P.~Huovinen,
%``Event-by-event distributions of azimuthal asymmetries in ultrarelativistic heavy-ion collisions,''
Phys. Rev. C \textbf{87}, 054901 (2013).
%doi:10.1103/PhysRevC.87.054901
arXiv:1212.1008 [nucl-th]

\bibitem{Sass:2025opk}
N.~Sass, H.~Roch, N.~G{\"o}tz, R.~Krupczak and C.~B.~Rosenkvist,
%``SPARKX: A Software Package for Analyzing Relativistic Kinematics in Collision Experiments,''
arXiv:2503.09415% [physics.data-an]].

\bibitem{Luzum:2012wu}
M.~Luzum and J.-Y.~Ollitrault,
%``Extracting the shear viscosity of the quark-gluon plasma from flow in ultra-central heavy-ion collisions,''
Nucl. Phys. A \textbf{904-905}, 377c (2013).
%doi:10.1016/j.nuclphysa.2013.02.028
arXiv:1210.6010 [nucl-th]

\bibitem{Gardim:2019brr}
F.~G.~Gardim, G.~Giacalone and J.-Y.~Ollitrault,
%``The mean transverse momentum of ultracentral heavy-ion collisions: A new probe of hydrodynamics,''
Phys. Lett. B \textbf{809}, 135749 (2020).
%doi:10.1016/j.physletb.2020.135749
arXiv:1909.11609 [nucl-th]

\end{thebibliography}

\end{document}